\documentclass[reprint,amssymb,noeprint,twocolumn,longbibliography]{revtex4-2}

\usepackage{hyperref}
\hypersetup{
colorlinks=true,
linkcolor=blue,
citecolor=blue,
filecolor=green,
urlcolor=blue,
}

\usepackage[utf8]{inputenc}
\usepackage{amsmath}
\usepackage{graphicx} 
\usepackage{physics}
\usepackage{float}
\usepackage[caption = false]{subfig}
\usepackage{color}
\usepackage{bm}

\usepackage{multirow}
\usepackage{epstopdf}
\usepackage[normalem]{ulem}

\usepackage{hyperref}
\hypersetup{
colorlinks=true,
linkcolor=blue,
citecolor=blue,
filecolor=green,
urlcolor=blue,
}
\usepackage[nameinlink,capitalise]{cleveref}

\begin{document} 

\title{Universality and threshold laws for collision lifetimes at ultralow temperatures}

\author{Rebekah Hermsmeier}
\affiliation{Department of Physics, University of Nevada, Reno, Nevada, 89557, USA}
\author{Timur V. Tscherbul}
\affiliation{Department of Physics, University of Nevada, Reno, Nevada, 89557, USA}

\date\today

\begin{abstract}
While collision lifetimes are a fundamental property of few-body scattering events, 
their behavior at ultralow temperatures is not completely understood.
We derive a general expression for the Smith lifetime Q-matrix using multichannel quantum defect theory, which allows us to obtain the   average time delay in the incident  $s$-wave collision channel in the limit of zero collision energy $E$, $Q_{11}=A/\sqrt{E} + B$, where $A$ and $B$ are constants.
We show that the time delay is dominated by elastic scattering, and contains an additional multichannel contribution independent of the two-body scattering length.
We also obtain the  expressions for the collision lifetime using the universal model of Idziaszek and Julienne [Phys. Rev. Lett. {\bf 104}, 113202 (2010)]. The lifetime acquires an imaginary part in the presence of inelastic loss due to the lack of unitarity of the S-matrix, and  shortens   with increasing the short-range loss parameter $y$.

 \end{abstract}
\maketitle
\newpage


\section{Introduction}
 
Ultracold few-body collisions and chemical reactions are responsible for a wide range of  phenomena, which occur in ultracold atomic and  molecular gases, such as thermalization, inelastic losses \cite{Bloch:08,Gross:17,Bohn:17}, and cluster formation \cite{Greene:17}. Particularly intriguing are   two-body collisions of ultracold molecules, which are characterized by the formation of long-lived collision complexes with  lifetimes  ranging from nanoseconds to milliseconds \cite{Bohn:17,Liu:22,Bause:23}.  These sticky collisions lead to inelastic losses, which are commonly described by the universal model (UM)  \cite{Idziaszek:10,Julienne:11} using a single parameter $y\in [0,1]$, which quantifies the extent of loss of the collision flux once the collision partners have reached the short-range region of the interaction potential. The value of the ultracold reaction rate in the universal limit ($y=1$)  does not depend on the details of short-range interactions.
Numerically exact coupled-channel calculations go beyond the UM and enable rigorous insights into short-range dynamics of the collision complex by providing the full S-matrix for a given interaction potential (see, e.g., Refs. \cite{Balakrishnan:16,Croft:17,Morita:18,Morita:19b,Park:23b,Karman:23,Morita:24fr}).


A complementary approach to characterizing bimolecular collisions is based on the concept of collision lifetime $Q=d\eta/dE$, where $\eta$ is the scattering phase shift and $E$ is the collision energy ($\hbar=1$ units are used throughout). As shown by Wigner and Eisenbud  
\cite{Wigner:55}, $Q$ gives the time delay experienced by a quantum wavepacket in the short-range interaction region as compared to the free wavepacket in the absence of interactions.
The concept of time delay was generalized by Smith \cite{Smith:60} to multichannel scattering.
The collision lifetime of a multichannel quantum scattering event is characterized in terms of the Q-matrix, whose diagonal elements give average time delays in each incident collision channel \cite{Smith:60}. 
The Q-matrix approach has proven fruitful for characterizing 
scattering resonances in atomic, molecular, and electron collisions \cite{FernandezAlonso:02,Wang:18,Kuppermann:95,Dunseath:00,Chao:03,Aquilanti:04,Aquilanti:05,Cavalli:07}. Importantly, the trace of the Q-matrix is related to the excess density of states of the collision complex (with respect to the unperturbed density of states) via the time delay theorem (see Ref.~\cite{Brumer:80} and references therein).
The zero-temperature limit of the time delay in single-channel elastic collisions is known to be $Q=-a\mu/k$ \cite{Field:03,Guillon:09,Simoni:09,Bovino:11,Croft:17b,Mehta:18,Frye:19}, where  $a$ is the scattering length, $\mu$ is the reduced mass of the collision partners, and $k$ is the collision wavevector.
(Alternative definitions of the time delay are also possible, such as $Q'=Q-\frac{1}{2E}\sin\eta$ \cite{Smith:60} with $Q'\to 0$ in the $s$-wave limit.)
However, ultracold collisions are typically multichannel in nature, i.e., they are accompanied by inelastic scattering, in which the internal states of the collision partners change. To our knowledge, the zero-temperature behavior of  multichannel  collision lifetimes is unknown.

Here, we use multichannel quantum defect theory (MQDT) to derive the threshold laws for collision lifetimes in multichannel scattering.
MQDT takes advantage of the separation of distance and energy scales in ultracold collisions to obtain their theoretical description in terms of
energy-independent short-range parameters.
MQDT is a powerful tool in the theory of ultracold atom-atom  \cite{Idziaszek:10,Julienne:11,Bohn:97,Gao:05,Gao:09,Gao:10,Mies:84,Mies:00,PerezRios:15} and atom-molecule \cite{Croft:11,Hazra:14} collisions. It has recently been combined with a frame transformation for greatly enhanced computational efficiency \cite{Morita:24mqdt}. Here, we obtain a general MQDT expression for the multichannel  Q-matrix and show that the average time delay in the incident collision channel (hereafter referred to as time delay or lifetime)  scales as $Q_{11}=A/k + B$, where $A$ and $B$ are constants, an analog of the Wigner threshold law for ultracold time delays. We also examine the role of elastic and inelastic contributions to the  time delay, and find that the former usually dominates.
Finally, we use the universal model of Idziaszek and Julienne \cite{Idziaszek:10,Julienne:11} to derive the universal limit for ultracold two-body collision lifetimes in the presence of inelastic losses.

\section{MQDT theory of collision lifetimes}

The lifetime Q-matrix may be expressed in terms of the scattering S-matrix as  \cite{Smith:60}
\begin{equation}
\begin{aligned}\label{Smith}
\textbf{Q}=i \textbf{S} \frac{d \textbf{S}^\dagger}{dE}.
 \end{aligned}
\end{equation}
The diagonal matrix elements of $\mathbf{Q}$, $Q_{ii}$, represent the average time delays in the $i$-th channel, with $i=1$ corresponding to the incident channel \cite{Smith:60}.
We begin with the MQDT expression for the multichannel S-matrix
\begin{equation}\label{label}
\begin{aligned}
\textbf{S}=e^{i \bm{\xi} }  [\textbf{I}+i \bar{\textbf{R}}][\textbf{I}-i \bar{\textbf{R}}]^{-1}e^{i\bm{\xi}},
\end{aligned}
\end{equation}
where $\mathbf{I}$ is the unit matrix, $\bm{\xi$} is a diagonal matrix containing the phase shifts due to long-range interactions, and $\bar{\textbf{R}}$ encapsulates short-range interactions \cite{Bohn:97,Gao:05,Gao:09,Gao:10,Mies:84,Mies:00,PerezRios:15,Croft:11,Hazra:14}. 
Combining this expression with Eq.~\eqref{Smith} we obtain the lifetime matrix as (see the Supplemental Material \cite{SM} for details of the derivation)
\begin{multline}\label{Qgen}
\textbf{Q}
=  \textbf{S} \bm{\xi}’ \textbf{S}^\dagger - \bm{\xi}’+  e^{i \bm{\xi}}  [\textbf{I}+i \bar{\textbf{R}}][\textbf{I}-i \bar{\textbf{R}}]^{-1}\\
\times
\bigl(\mathbf{I}+[\textbf{I}-i \bar{\textbf{R}}] [{\mathbf{I}} +i \bar{\textbf{R}}]^{-1}\bigr)  \mathbf{\bar{R}}' [\mathbf{I}+i \bar{\textbf{R}}]^{-1}  e^{-i\bm{\xi}},
\end{multline}
where $\mathbf{\bar{R}}'=d\bar{\textbf{R}}/dE$ and
\begin{equation}\label{label2}
\begin{aligned}
\bar{\textbf{R}}=\textbf{C}^{-1}[\bar{\textbf{Y}}^{-1}-\tan{\bm{\lambda}})]^{-1}\textbf{C}^{-1}.\\
 \end{aligned}
\end{equation}
Here,  the diagonal matrices $\tan{\bm{\lambda}}$ and $\textbf{C}^{-1}$ account for  threshold effects and the relationship between the amplitude of the WKB-normalized and energy-normalized reference functions \cite{Idziaszek:10}.
The matrix  $\bar{\textbf{Y}}$ is expressed via the open-open (oo), open-closed (oc), and closed-closed (cc) subblocks of yet another matrix $\mathbf{Y}$ as  $\bar{\textbf{Y}}=\textbf{Y}_{oo}-\textbf{Y}_{oc}[\tan\bm{\nu} +\textbf{Y}_{cc}]^{-1} \textbf{Y}_{co}$, where $\tan\bm{\nu}$ is a diagonal matrix of closed-channel phases. The elements of 
$\textbf{Y}$ can be assumed independent of the collision energy $E$. The matrix  $\bar{\textbf{Y}}$ contains 
closed-channel effects (which give rise to, e.g.,  Feshbach   resonances) and can therefore vary rapidly with $E$.
Here, we are interested in threshold effects on multichannel collision lifetimes, so we will neglect these variations  and set $\bar{\textbf{Y}}=\textbf{Y}_{oo}$, which is equivalent to assuming that we are far away from any resonances. 
The term $\textbf{Y}_\lambda=[\textbf{Y}_{oo}^{-1}-\tan(\bm{\lambda})]^{-1}$ is energy-independent because $\tan\bm{\lambda}$ is energy-independent near $s$ and $p$-wave thresholds \cite{Idziaszek:10}.
As a result, Eq.~\eqref{label2} becomes
$ \bar{\textbf{R}}=\textbf{C}^{-1} \textbf{Y}_\lambda \textbf{C}^{-1}$, so $\mathbf{\bar{R}}'=0$ and the last term in Eq.~\eqref{Qgen} vanishes.

We are interested in evaluating the average time delay in the incident collision channel ($Q_{11}$).
To this end, we assume that the energy gap between the incident channel and all other inelastic channels is large compared to  the collision energy.
This allows us to neglect the energy dependence of all but the leading MQDT parameters in the diagonal matrices $\tan\bm{\lambda}$ and $\mathbf{C}$.
To isolate the energy dependent terms, we decompose the $N\times N$ $\bar{\textbf{R}}$ matrix into $1\times 1$ and $(N-1)\times (N-1)$ blocks and use the Schur block inversion method to calculate $[\textbf{I}-i \bar{\textbf{R}}]^{-1}$ in Eq.~\eqref{Qgen} (see the Supplemental Material \cite{SM}). This procedure gives the elastic S-matrix element as (we highlight energy-dependent MQDT terms in magenta)
 \begin{equation}
\begin{aligned}\label{S11}
 S_{11}=e^{2 i \textcolor{magenta}{\xi_1}} \frac{1+ {a_1}{\textcolor{magenta}{C_1^{-2}}} }{1+{\textcolor{magenta}{C_1^{-2}}} M_N},  
\end{aligned}
\end{equation}
where 
\begin{equation}
\begin{aligned}\label{consta}
  a_1 = i Y_{\lambda_{11}} +
i \frac{Y_{\lambda_{12}}}{C_2} (i \tilde{D}_{11} \frac{Y_{\lambda_{12}}}{C_2} +i \tilde{D}_{12} \frac{Y_{\lambda_{13}}}{C_3}+...)\\
+i\frac{Y_{\lambda_{13}}}{C_3} (i \tilde{D}_{12} \frac{Y_{\lambda_{12}}}{C_2}+i \tilde{D}_{22} \frac{Y_{\lambda_{13}}}{C_3}+...)+ ... \\
\\
M_N=-iY_{\lambda_{11}}+
\frac{1}{C_2}Y_{\lambda_{12}}
(\frac{\tilde{D}_{11}}{ C_2} Y_{\lambda_{12}} +\frac{\tilde{D}_{12}}{ C_3} Y_{\lambda_{13}} +... )\\
+\frac{1}{ C_3} Y_{\lambda_{13}}
(\frac{\tilde{D}_{21}}{ C_2} Y_{\lambda_{12}} +\frac{\tilde{D}_{22}}{ C_3} Y_{\lambda_{13}}+...)+...\, ,
 \end{aligned}
\end{equation}
are energy-independent complex constants, and 
the matrix elements $\tilde{D}_{ij}$ are defined in the Supplemental Material \cite{SM}.
The first column of the S-matrix is given by 
\begin{equation}\label{Sfirst_col}
\begin{aligned}
S_{n1}=e^{i\textcolor{magenta}{\xi_1}} e^{i\xi_n} 
\frac{C_1^{-1}}{1+{\textcolor{magenta}{C_1^{-2}}} M_N} b_n,
 \end{aligned}
\end{equation}
where $n\ge 2$
and
\begin{equation}
\begin{aligned}\label{constb}
  b_n= i \frac{Y_{\lambda_{1n}}}{C_n} +(i\tilde{D}_{1(n-1)} \frac{Y_{\lambda_{12}}}{C_2}+i\tilde{D}_{2(n-1)} \frac{Y_{\lambda_{13}}}{C_3}+...)\\
  +i \frac{Y_{\lambda_{2n}}}{C_2 C_n}  (i \tilde{D}_{11} \frac{Y_{\lambda_{12}}}{C_2} +i \tilde{D}_{12} \frac{Y_{\lambda_{13}}}{C_3}+...)\\
  +  i \frac{Y_{\lambda_{3n}}}{C_3 C_n} (i \tilde{D}_{12} \frac{Y_{\lambda_{12}}}{C_2}+i \tilde{D}_{22} \frac{Y_{\lambda_{13}}}{C_3}+...)
+...
 \end{aligned}
\end{equation}
The energy dependence of the rest of the S-matrix is derived in the Supplemental Material \cite{SM}. 

To obtain the average time delay ${Q}_{11}$, we also require the energy derivative of the first column of the S-matrix, which is readily obtained from Eqs.~\eqref{S11} and \eqref{Sfirst_col} as
 \begin{equation}
\begin{aligned} 
\frac{d S_{11}}{d{E}}=2 i \textcolor{magenta}{\xi_1'} e^{2 i \textcolor{magenta}{\xi_1}} \frac{1+ {a_1}{\textcolor{magenta}{C_1^{-2}}}}{1+{\textcolor{magenta}{C_1^{-2}}} M_N}  +e^{2 i \textcolor{magenta}{\xi_1}} \frac{ 2 \textcolor{magenta}{C_1}  \textcolor{magenta}{C_1^\prime} (M_N-a_1) }{(\textcolor{magenta}{C_1^2} + M_N)^2} 
\end{aligned}
\end{equation}
and
\begin{equation}\label{Sn1der}
\begin{aligned}
\frac{d S_{n1}}{d{E}}
= e^{i \textcolor{magenta}{\xi_1}} e^{i\xi_n} b_n \left[  \frac{i \textcolor{magenta}{\xi_1^\prime} {\textcolor{magenta}{C_1^{-1}}}}{1+{\textcolor{magenta}{C_1^{-2}}} M_N} + 
\frac{\textcolor{magenta}{C_1}^\prime (M_N-\textcolor{magenta}{C_1^2})}{(\textcolor{magenta}{C_1^2}+ M_N)^2} \right]
 \end{aligned}
\end{equation}
for $n\ge 2$.
Using the above results in combination with Eq.~\eqref{Smith}, expressing $M_N=M_R + iM_I$, and noting that $4 M_R=\sum_n |b_n|^2$, we find   \cite{SM}
\begin{equation}
\begin{aligned}\label{q1111}
Q_{11}=  2  \textcolor{magenta}{\xi_1^\prime} \frac{|1+ {a_1}{\textcolor{magenta}{C_1^{-2}}}|^2}{|1+{\textcolor{magenta}{C_1^{-2}}} M_N|^2}  + \frac{ 4 \textcolor{magenta}{C_1}  \textcolor{magenta}{C_1^\prime}(M_I+{\textcolor{magenta}{C_1^{-4}}}(M_R^2 M_I+M_I^3))}{|1+{\textcolor{magenta}{C_1^{-2}}} M_N^*|^2 |\textcolor{magenta}{C_1^2}+ M_N|^2} \\
+  \frac{\textcolor{magenta}{\xi_1^\prime}}{\textcolor{magenta}{C_1}^2}
\frac{4 M_R}{|1+\frac{1}{\textcolor{magenta}{C_1^2}} M_N|^2} + \frac{8 \textcolor{magenta}{C_1^\prime} M_R M_I  }{\textcolor{magenta}{C_1}  |\textcolor{magenta}{C_1^2}+ M_N|^2 |1+  M_N^*{\textcolor{magenta}{C_1^{-2}}}|^2 } ,\\
 \end{aligned}
\end{equation}
This equation gives the average time delay in the incident collision channel in terms of energy-dependent MQDT parameters, a key result of this work.

To proceed, we need to specify the energy dependence of the MQDT  parameters. We focus here on the s-wave threshold regime ($E\to0$ and $l=0$), where  \cite{Idziaszek:10} 
$C^{-2}=k\bar{a}[1+(s-1)^2]$, $\tan\xi=-ka$, and $\tan\lambda=1-s$. 
Here, 
$s={a}/{\bar{a}}$ is the dimensionless scattering length, and $\bar{a}=4 \pi C_6/\Gamma(\frac{1}{4})^2$ is the mean scattering length expressed in terms of the long-range dispersion coefficient $C_6$ \cite{Idziaszek:10}. Substituting these values in Eq.~\eqref{q1111} we find
\begin{equation}\label{q11final}
\begin{aligned}
Q_{11}=
  -\frac{2 a {\mu} }{ \textcolor{magenta}{k}(1+   a^2 \textcolor{magenta}{k^2})} \frac{1-2M_R \textcolor{magenta}{k }\bar{a} [1+(s-1)^2]}{1+2M_R \textcolor{magenta}{k }\bar{a} [1+(s-1)^2]} 
\\
-   2  {\mu}\frac{1}{  \textcolor{magenta}{k}} \frac{M_I(\bar{a} [1+(s-1)^2] )}{1+4M_R \textcolor{magenta}{k }\bar{a} [1+(s-1)^2]} \\
-  \frac{a\mu }{1+   a^2 \textcolor{magenta}{k^2}} \frac{4 M_R
\bar{a} [1+(s-1)^2] }{1+2M_R \textcolor{magenta}{k }\bar{a} [1+(s-1)^2]} \\
- 4 {\mu} \frac{M_R M_I  \bar{a}^{2} [1+(s-1)^2]^{2} }{1+4M_R \textcolor{magenta}{k }\bar{a} [1+(s-1)^2]}.
 \end{aligned}
\end{equation}
We observe that the  energy dependence of the average time delay is determined by four contributions.
The first two contributions dominate in the limit $k\to 0$, where the first term reduces to $-2a\mu/k$, in agreement with prior single-channel results \cite{Field:03,Guillon:09,Simoni:09,Bovino:11,Croft:17b,Mehta:18,Frye:19}.
The second term in Eq.~\eqref{q11final}
gives an intrinsically multichannel  contribution to the average time delay, which vanishes for single-channel elastic scattering. 
The third and the fourth terms are also of multichannel origin, but approach constant values in the limit $k\to 0$.
We analyze these contributions in more detail below.

  \begin{figure}[t]
\begin{center}
\subfloat[$Y_{12}$=1000]{\includegraphics[width = 1.5in]{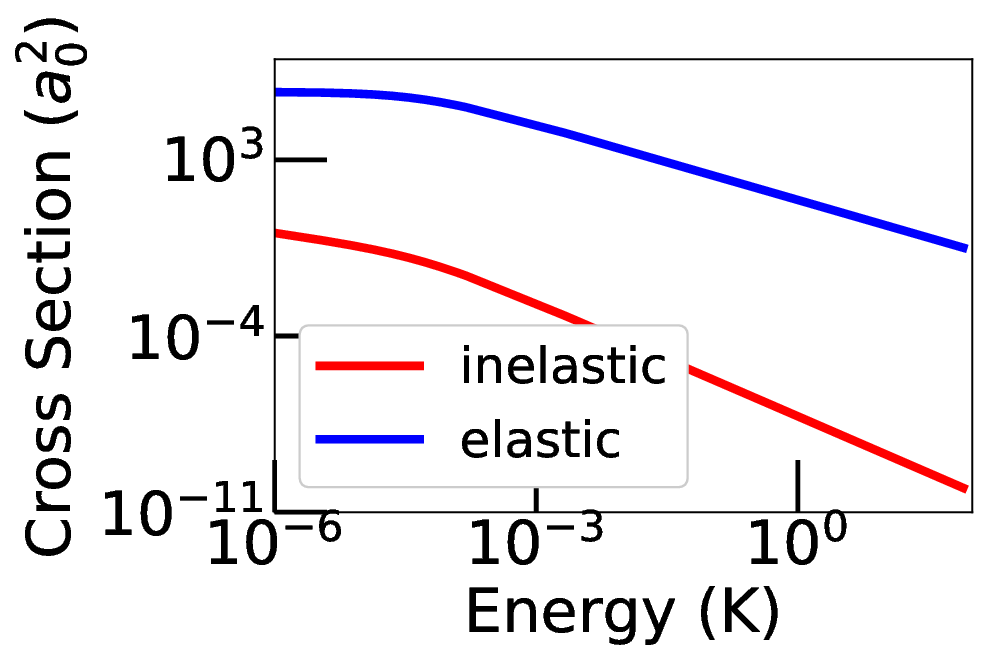}} 
\subfloat[$Y_{12}$=10]{\includegraphics[width = 1.5in]{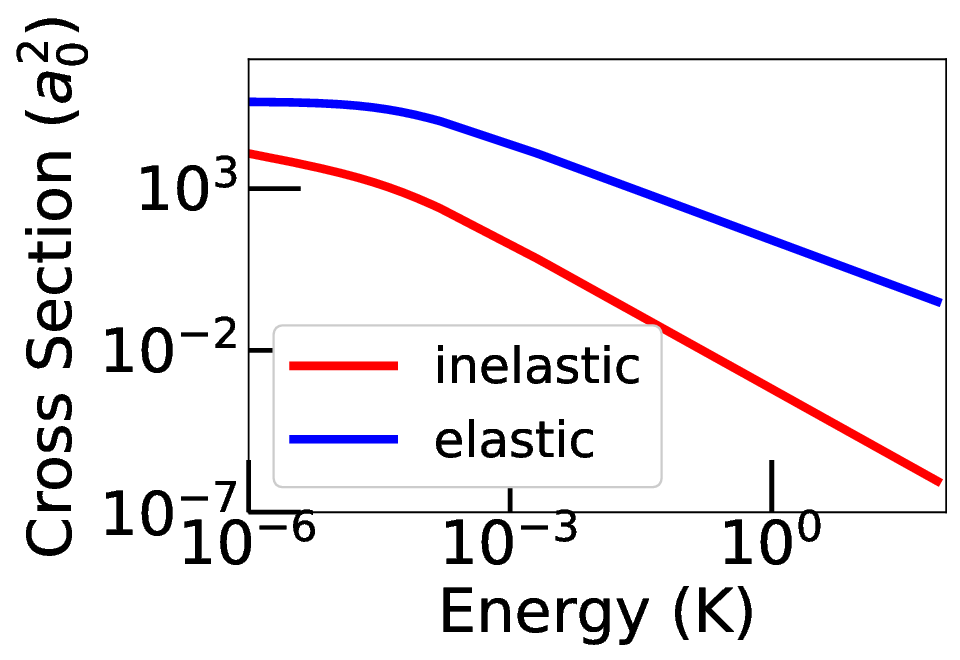}}\\
\subfloat[$Y_{12}$=0.1]{\includegraphics[width = 1.5in]{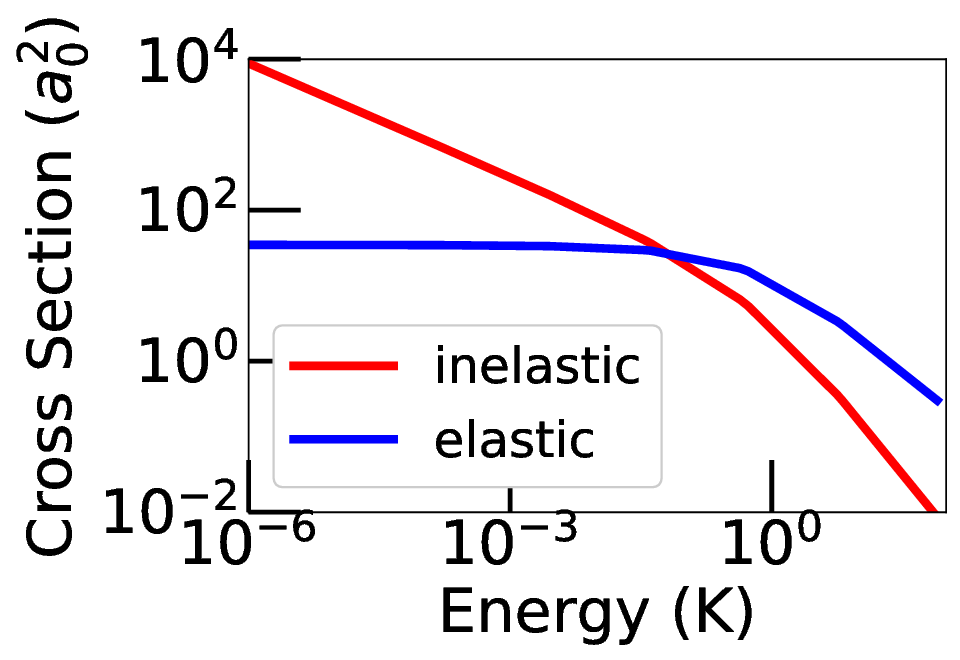}}
\subfloat[$Y_{12}$=0.0001]{\includegraphics[width = 1.5in]{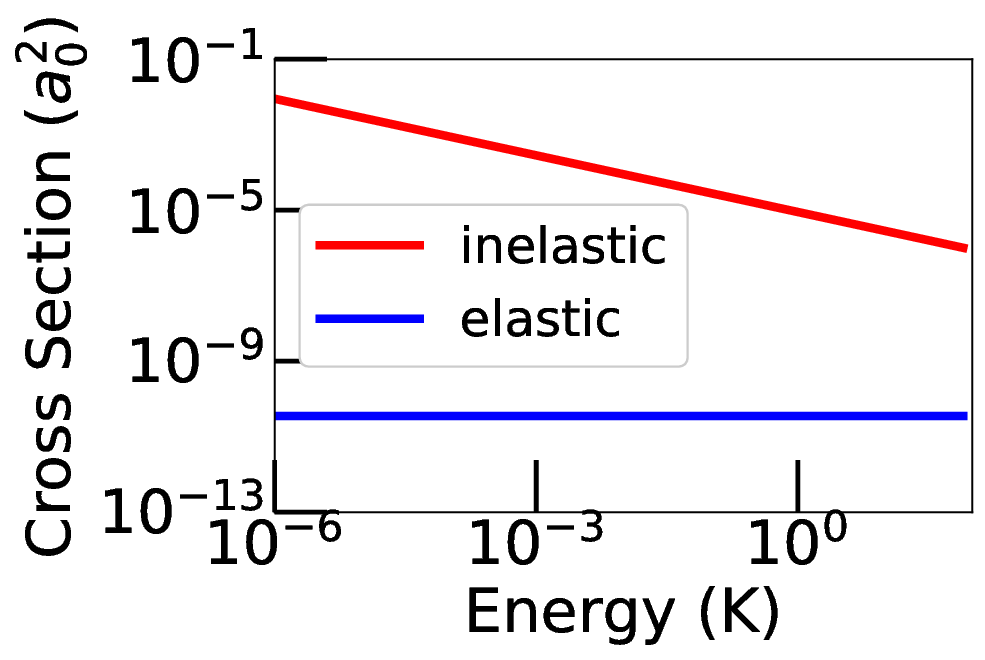}} 
\caption{Elastic and inelastic cross sections as a function of collision energy in the two-channel model.}
\end{center}
\label{Y12elastic}
\end{figure}


To gain more insight into the threshold scaling of the average time delay, consider a two-channel model described by a single short-range parameter $Y_{12}$ \cite{Jachymski:14,Chilcott:22}. The  choice $Y_{11}=Y_{22}=0$ simplifies the calculations without loss of generality \cite{SM}.
The corresponding elastic and inelastic  cross sections are plotted as a function of collision energy in Fig.~1.
At low energies, the cross sections follow the expected Wigner threshold laws $\sigma(k) \simeq \text{const}$ ($\sigma(k) \simeq {1}/{k}$) for s-wave elastic (inelastic) scattering.


  \begin{figure}[t]
\begin{center}
\subfloat[$Y_{12}$=1000]{\includegraphics[width = 1.5in]{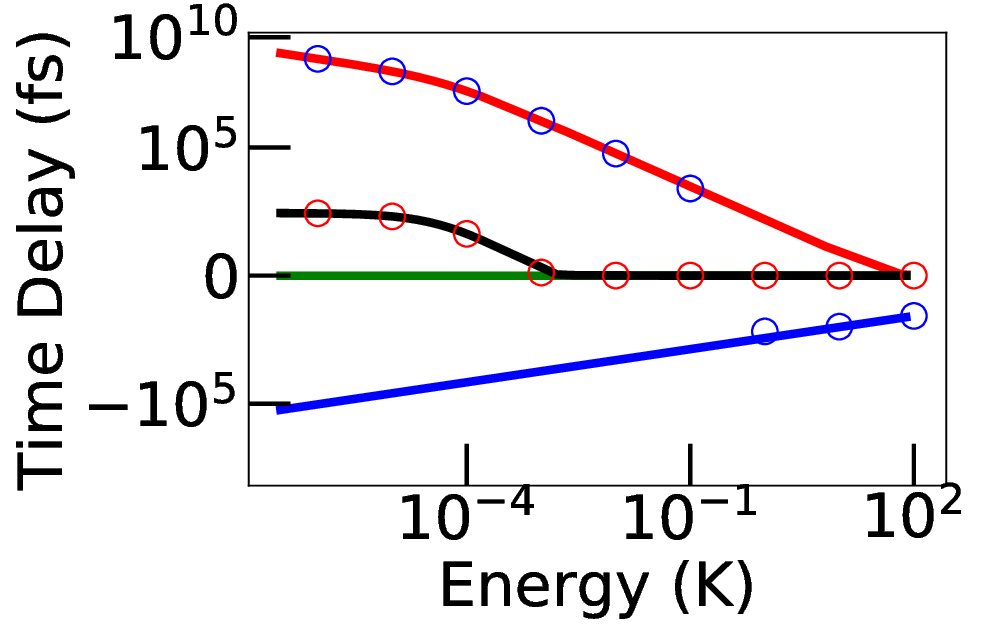}} 
\subfloat[$Y_{12}$=10]{\includegraphics[width = 1.5in]{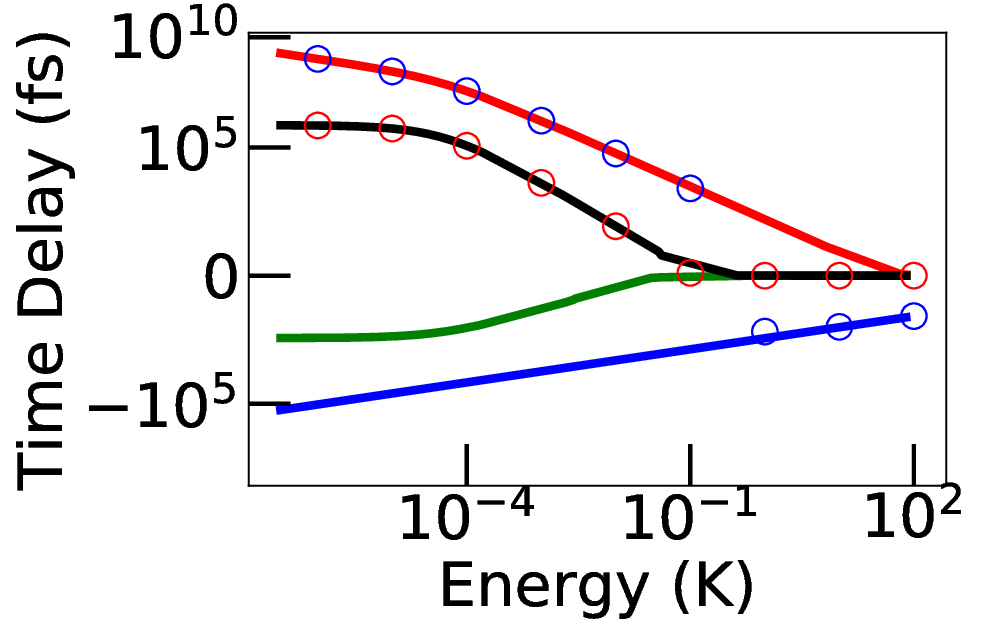}}\\
\subfloat[$Y_{12}$=0.1]{\includegraphics[width = 1.5in]{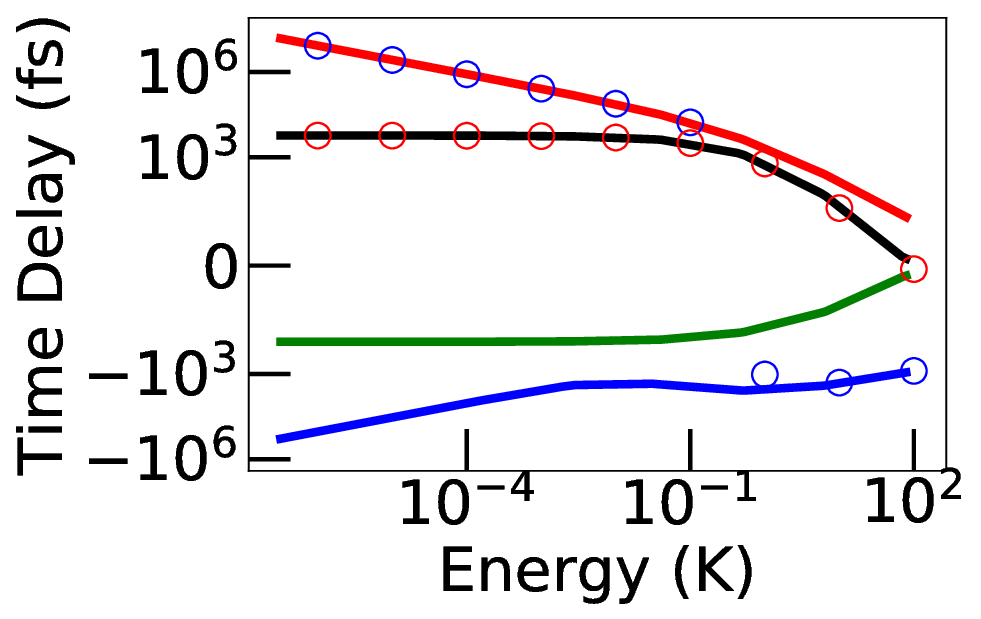}}
\subfloat[$Y_{12}$=0.0001]{\includegraphics[width = 1.5in]{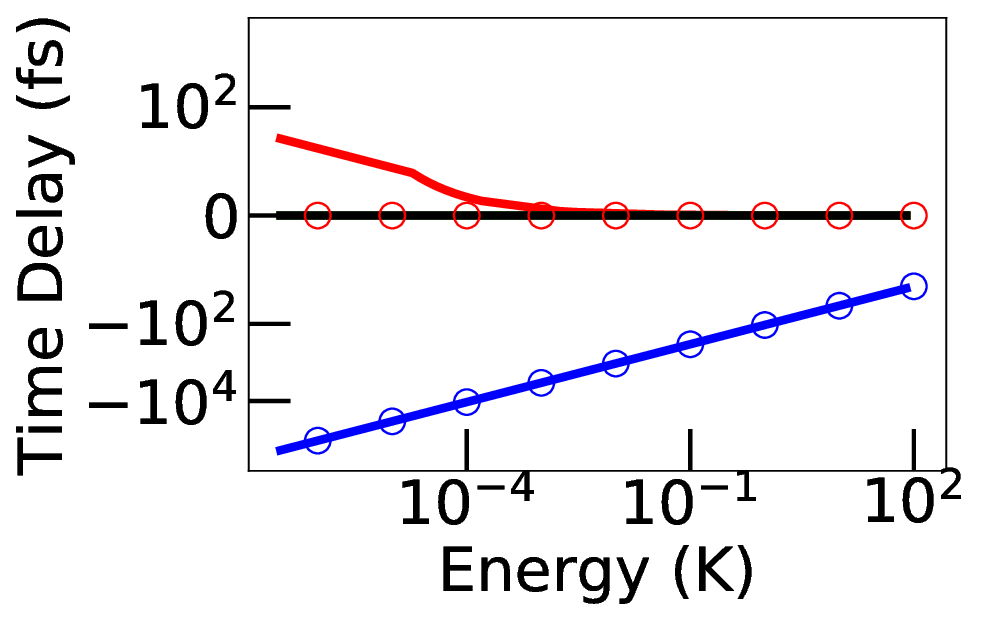}} 
\caption{ The four contributions to the average time delay $Q_{11}$ in the two-channel model [see Eq.~\eqref{q11final}] plotted as a function of collision energy. 
{Solid blue and red lines show the elastic contributions [the first two terms in Eq.~\eqref{q11final}].   Solid green and black lines show the inelastic contributions  [the last two terms in Eq.~\eqref{q11final}]. Blue circles -- the total $Q_{11}$, red circles -- the total inelastic contribution to $Q_{11}$.}}
\end{center}
\label{Y12Qpart}
\end{figure}

While one can derive closed-form analytic expressions for the average time delay in the two-channel model \cite{SM}, they are too cumbersome to analyze.
Instead, we plot in Fig. 2 the four contributions to $Q_{11}$ [see Eq.~\eqref{q11final}] as a function of collision energy 
for different values of $Y_{12}$. 
As noted above, the first and the second terms  in Eq.~\eqref{q11final} arise from elastic collisions, while the third and the forth terms  arise from inelastic collisions. We observe that the elastic contributions to the average time delay dominate at all energies and $Y_{12}$ values.
This holds 
regardless of the value of $Y_{12}$, and 
is further emphasized in Fig.~2, where the total inelastic contribution is compared with the total value of $Q_{11}$ and found to be significantly smaller in comparison. This is consistent with Eq.~\eqref{q11final}, which shows that in the $k\to 0$ limit, the elastic contributions scale as $Q_{11}^\text{el}(k) \simeq {1}/{k}$, whereas the inelastic ones remain constant, so the former  are  expected to dominate.

Physically, the predominance of the elastic scattering contribution to the time delay can be understood by noting that, in the limit of zero collision energy, the inelastically scattered wavepacket  escapes from the collision region with a much higher velocity than either the incident or the elastically scattered wavepacket. As a result, the lifetime of the inelastically scattered wavepacket  makes a negligible contribution  to the average time delay, consistent with the threshold law in Eq.~\eqref{q11final}.




\section{Universal model of collision lifetimes}

The $S$-matrix expression in the UM takes the form \cite{Idziaszek:10}
\begin{equation}\label{Smat_UM}
\begin{aligned}
S_{11}= \frac{1+i (\tan\xi(E)-\frac{y C^{-2}(E)}{i+y\tan\lambda(E)})}{1-i  (\tan\xi(E)-\frac{y C^{-2}(E)}{i+y\tan\lambda(E)})},
\end{aligned}
\end{equation}
where $\xi(E)$, $C(E)$, and $\tan\lambda(E)$ are MQDT parameters (see above)   \cite{Gao:98,Idziaszek:10} and $y$ is an energy-independent parameter, which quantifies the loss of the short-range collision flux ($y=0$  corresponds to no loss, and $y=1$ to complete loss, also known as the  universal limit \cite{Idziaszek:10}).

Taking the energy derivative of Eq.~\eqref{Smat_UM} we obtain for the average collision time delay $Q_{11}=- i S_{11}^\dagger \frac{d S_{11}}{dE}$ \cite{SM}
\begin{equation}\label{Q11_UM}
\begin{aligned}
-i   \frac{1-i (\tan\xi-\frac{y C^{-2}(E)}{-i+y\tan\lambda})}{1+i  (\tan\xi-\frac{y C^{-2}(E)}{-i+y\tan\lambda})}  \frac{ 2 i (\frac{d\tan\xi}{dE} + 2 \frac{y}{-i+y\tan\lambda} \frac{C’(E)}{C^{3}(E)})}{(1-i  (\tan\xi-\frac{y C^{-2}(E)}{i+y\tan\lambda}))^2}
\end{aligned}
\end{equation}
Using the values of MQDT functions   in the $s$-wave limit \cite{Idziaszek:10}
 Eq.~\eqref{Q11_UM} becomes $Q_{11}=Q_{11}^R+iQ_{11}^I$ with \cite{SM}
\begin{align}\label{Q11UM}\notag
Q_{11}^R &= \frac{2\mu}{k}[aF_1(s,y) + y\bar{a}(2-s)F_2(s,y)], \\
Q_{11}^I &= \frac{2\mu}{k}[aF_2(s,y) - y\bar{a}(2-s)F_1(s,y)], 
\end{align}
where $F_1(s,y)=(-2s^2y^2 + 4sy^2 -2y^2 - 1)/D$, $F_2(s,y)=(s^3y^3 - 3s^2y^3 + 3sy^3-sy-y^3+y)/D$, and $D=[1+y^2(1-s)^2]^2$ are  auxiliary functions.

\begin{figure}[t!]
\begin{center}
\subfloat[$s$=1]{\includegraphics[width = 1.5in]{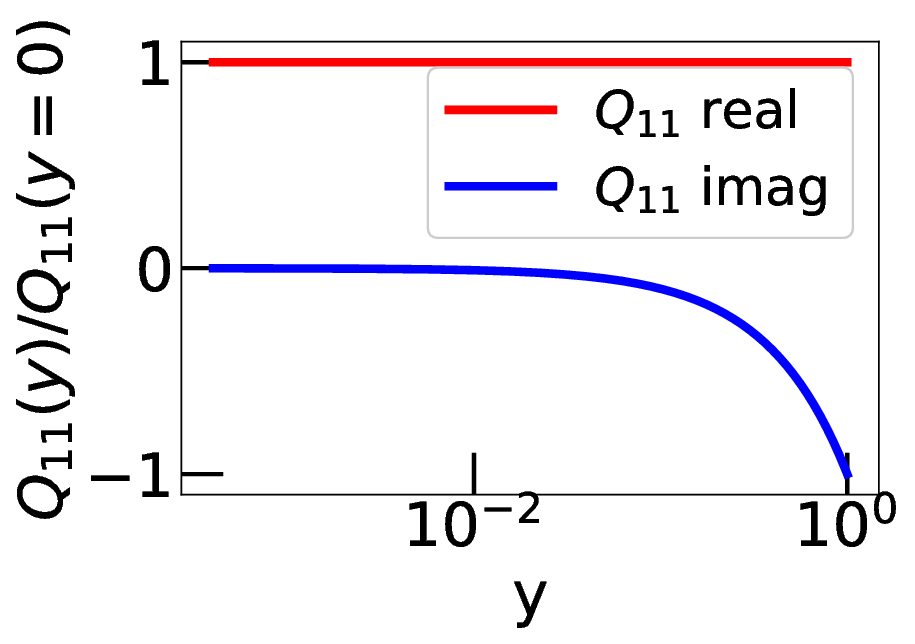}} 
\subfloat[$s$=2]{\includegraphics[width = 1.5in]{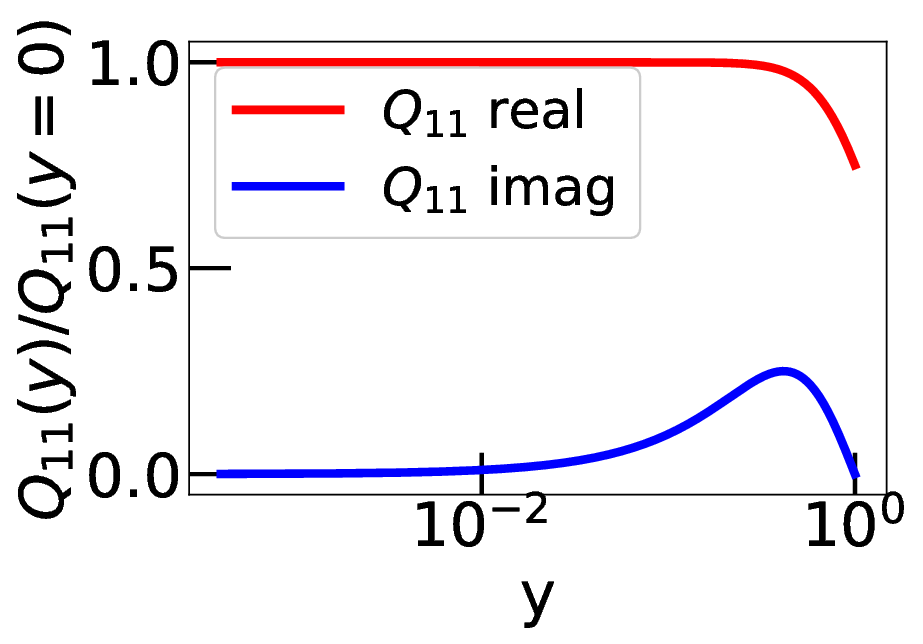}}\\
\vspace{-0.2cm}
\subfloat[$s$=10]{\includegraphics[width = 1.5in]{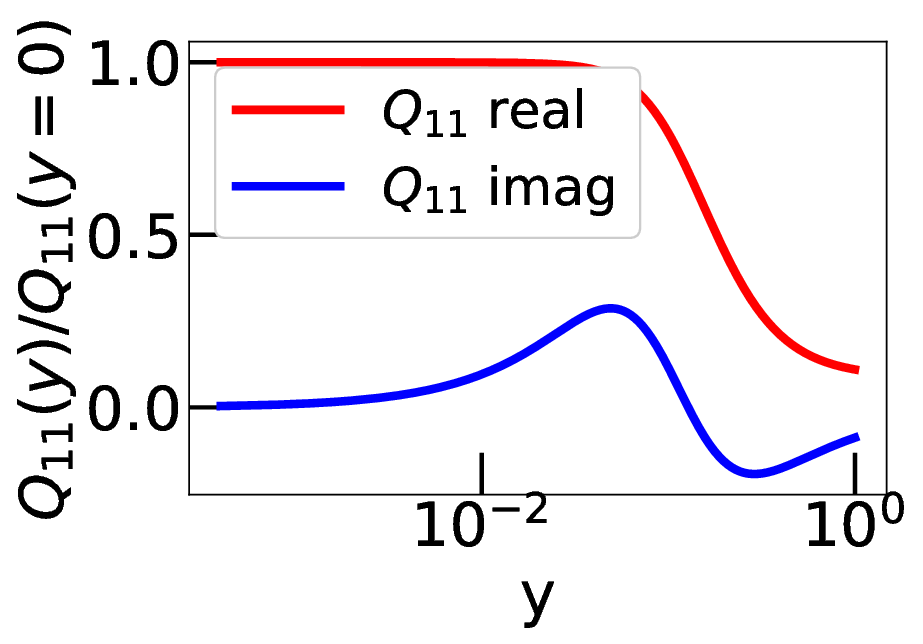}}
\subfloat[$s$=100]{\includegraphics[width = 1.5in]{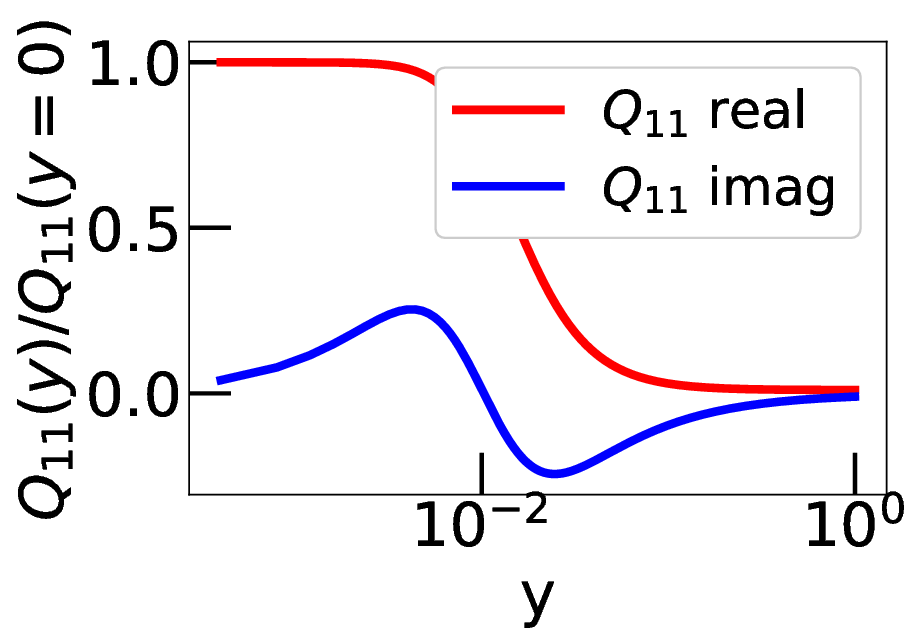}} 
\caption{The ratio of $s$-wave time delays $Q_{11}/Q_{11}(y=0)$ given by Eq.~\eqref{Q11UM} as a function of the short-range loss parameter $y$ for different values of the reduced scattering length $s$. Note that the ratio is independent of $k$.}
\label{ysQpart}
\end{center}
\end{figure}

The time delay given by Eq.~\eqref{Q11UM} is a complex quantity because the scattering matrix in the UM \eqref{Smat_UM} is unitary only
in the limit $y\to 0$ (no loss), where $F_1(s,y)\to -1$ and $F_2(s,y)\to 0$, and the expressions \eqref{Q11UM} reduce to the prior result obtained for one-dimensional potential scattering $Q_{11}= -2\mu a/k$  \cite{Field:03,Guillon:09,Simoni:09,Bovino:11,Croft:17b,Mehta:18,Frye:19}.
Thus, the real part of the complex time delay ($Q_{11}^R$) can be interpreted as arising from elastic scattering, and the imaginary part ($Q_{11}^R$) as due to inelastic scattering. This is similar to  the imaginary part of the scattering length \cite{Balakrishnan:97,Yang:10,Idziaszek:10,Hutson:07}, which is proportional to the inelastic scattering cross section.
The universal limit for the average time delay may be obtained by setting $y\to 1$ in Eq.~\eqref{Q11UM}.

Figure \ref{ysQpart} shows the dependence of the average time delay (normalized by its value in the absence of inelastic loss) on the loss parameter $y$ for different values of $s$.
For small and moderate $s\leq 10$, the real (elastic) part of the time delay  always dominates. The imaginary contribution  $Q_{11}^I$ is very small for small $y$, but increases with $y$, which is consistent with the interpretation of $Q_{11}^I$ as due to inelastic loss. Interestingly, we observe a single maximum for $s=2$ (and a maximum followed by a minimum for $s\ge 10$) in the imaginary time delay as a function of $y$. These maxima and minima occur near $y=1$ for small $s$ but shift to smaller $y$ for larger $s$. We attribute these variations to the quantum interference between the flux partially reflected from the short-range region (for $y<1$) and the incident flux \cite{Idziaszek:10,Morita:23}.

Interestingly, as shown in Fig.~3, the ratio $Q_{11}^R(y)/Q_{11}(y=0)<1$  is smaller than unity for all values of $s$ and $y$ studied. Thus, the average time delay in the presence of inelastic loss ($y>1$) tends to be smaller in absolute magnitude than the single-channel time delay $|Q_{11}(y=0)| = 2\mu |a|/k$. The latter may therefore be regarded as an upper limit to the time delay that can be reached in a multichannel collision (in the absence of resonant scattering).

\vspace{-0.35cm}
\section{Summary}

In summary, we have developed the MQDT  theory of collision lifetimes based on Smith's Q-matrix formalism.  
This enabled us to  derive the Wigner threshold law for multichannel collision lifetimes, extending previous single-channel results \cite{Field:03,Guillon:09,Simoni:09,Bovino:11,Croft:17b,Mehta:18,Frye:19}.  We found that
the zero-energy behavior of the average time delay in a simple two-channel MQDT model is dominated by elastic scattering in most regimes. 
Finally, we have  extended  the UM of lossy collisions \cite{Idziaszek:10,Julienne:11}  to calculate the collision lifetimes, and explored their dependence on the short-range loss parameter $y$, again finding the predominance of elastic scattering in all regimes, except at large $y\simeq 1$ and  $s\gg 10$.
In future work, it would be interesting to investigate multichannel collision lifetimes in the presence of closed channels, which give rise to Feshbach resonances.

\section*{Acknowledgements}
This work was supported by the NSF CAREER program (grant No. PHY-2045681).

\newpage

\widetext
\begin{center}
\textbf{\large Supplemental Material}

\end{center}
\setcounter{equation}{0}
\setcounter{figure}{0}
\setcounter{table}{0}
\setcounter{section}{0}

\tableofcontents

\vspace{0.8cm}
In this Supplemental Material (SM) we provide a derivation of the expressions for the lifetime Q-matrix and  the average time delay $Q_{11}$ using  multichannel quantum defect theory (MQDT) [see Secs.~\ref{sec:SM_PESs} through \ref{two_channel_model}]. 
Section~ \ref{sec:Q11UM} describes the derivation of $Q_{11}$ in the universal model (UM) of Ref.~\cite{Idziaszek:10}.


 

\setcounter{equation}{0}
\setcounter{figure}{0}

\vspace{0.5cm}
\section{\label{sec:SM_PESs}{Scattering and lifetime matrices: Multichannel quantum defect theory expressions}}

The lifetime matrix $\textbf{Q}$ is defined as \cite{Smith:60} 
\begin{equation}
\begin{aligned}\label{Smith}
\textbf{Q}=i  \textbf{S} \frac{d \textbf{S}^\dagger}{dE},
 \end{aligned}
\end{equation}
The diagonal matrix elements $Q_{ii}$ of $\textbf{Q}$ represent average time delays in the $i$-th channel. 
To evaluate $\mathbf{Q}$, we begin with the standard MQDT expression for the scattering $S$-matrix \cite{Mies:84,Mies:00,Gao:05,Croft:11,Hazra:14}
\begin{equation}\label{label}
\begin{aligned}
\textbf{S}=e^{i \bm{\xi}}  [\textbf{I}+i \bar{\textbf{R}}][\textbf{I}-i \bar{\textbf{R}}]^{-1}e^{i\bm{\xi}}
\end{aligned}
\end{equation}
Here, $\bm{\textbf{$\xi$}}$ is a diagonal matrix containing the phase shifts due to long-range interactions, and the matrix $\bar{\textbf{R}}$ (defined below) encapsulates short-range interactions. 
Using this definition, the lifetime matrix $\textbf{Q}$ is given by (in atomic units, where $\hbar=1$)
\begin{equation}
\begin{aligned}
\textbf{Q}=i  \textbf{S} \frac{d \textbf{S}^\dagger}{dE}
&=   \textbf{S} \bm{\xi}’ \textbf{S}^\dagger - \bm{\xi}’+  \textbf{S} e^{-i \textbf{$\xi$}} \bigg(1+[\textbf{I}-i \bar{\textbf{R}}] [1+i \bar{\textbf{R}}]^{-1}\bigg)  \bar{\textbf{R}}^\prime [1+i \bar{\textbf{R}}]^{-1}  e^{-i\bm{\xi}} \\
&=    \textbf{S} \xi’ \textbf{S}^\dagger - \bm{\xi}’+  e^{i \bm{\xi}}  [\textbf{I}+i \bar{\textbf{R}}][\textbf{I}-i \bar{\textbf{R}}]^{-1}\bigg(1+[\textbf{I}-i \bar{\textbf{R}}] [1+i \bar{\textbf{R}}]^{-1}\bigg)  \bar{\textbf{R}}^\prime [1+i \bar{\textbf{R}}]^{-1}  e^{-i\bm{\xi}},
 \end{aligned}
\end{equation}
where $\bar{\textbf{R}}'=d\bar{\textbf{R}}/dE$ with the $\bar{\textbf{R}}$ matrix defined as  \cite{Croft:11}
\begin{equation}\label{label2}
\begin{aligned}
\bar{\textbf{R}}=\textbf{C}^{-1}[\bar{\textbf{Y}}^{-1}-\tan(\bm{\lambda})]^{-1}\textbf{C}^{-1},\\
 \end{aligned}
\end{equation}
where the diagonal matrices $\tan({\bm{\lambda}})$ and $\textbf{C}^{-1}$ account for  threshold effects and matching between the WKB-normalized and energy-normalized reference functions \cite{Mies:84,Mies:00,Gao:05,Croft:11}.
The matrix  $\bar{\textbf{Y}}$ is expressed via the open-open (oo), open-closed (oc), and closed-closed (cc) subblocks of an energy-independent matrix $\mathbf{Y}$ as  $\bar{\textbf{Y}}=\textbf{Y}_{oo}-\textbf{Y}_{oc}[\tan\bm{\nu} +\textbf{Y}_{cc}]^{-1} \textbf{Y}_{co}$, where $\tan\bm{\nu}$ is a diagonal matrix of closed-channel phases.   As noted in the main text, we assume that closed-channel effects can be neglected,  so that  the term   $ [\textbf{Y}_{oo}^{-1}-\tan(\lambda)]^{-1}=\textbf{Y}_\lambda$ is  energy-independent.  We can thus write
\begin{equation}
\begin{aligned}
\bar{\textbf{R}}=\textbf{C}^{-1}[\textbf{Y}_{oo}^{-1}-\tan(\lambda)]^{-1}\textbf{C}^{-1} =  \textbf{C}^{-1} \textbf{Y}_\lambda \textbf{C}^{-1}.
 \end{aligned}
\end{equation}



 Using matrix notation
 \begin{equation}
\begin{aligned}
 \textbf{C}^{-1}
=\begin{pmatrix} 
\frac{1}{C_1}&0&0&...\\
0&\frac{1}{C_2}&0&...\\
0&0&\frac{1}{C_3}&...\\
...&...&...&...
\end{pmatrix}, \quad
\textbf{Y}_\lambda  =
\begin{pmatrix} 
\frac{1}{C_1}&0&0&...\\
0&\frac{1}{C_2}&0&...\\
0&0&\frac{1}{C_3}&...\\
...&...&...&...
\end{pmatrix},\\
\end{aligned}
\end{equation}
which gives
\begin{equation}
\begin{aligned}
\bar{\textbf{R}} = \begin{pmatrix} 
\frac{1}{C_1^2}Y_{\lambda_{11}}&\frac{1}{C_1 C_2}Y_{\lambda_{12}} &\frac{1}{C_1 C_3} Y_{\lambda_{13}} &...\\
\frac{1}{C_1 C_2} Y_{\lambda_{12}}&\frac{1}{C_2^2} Y_{\lambda_{22}}&\frac{1}{C_2 C_3} Y_{\lambda_{23}}&...\\
\frac{1}{C_1 C_3} Y_{\lambda_{13}}&\frac{1}{C_2 C_3} Y_{\lambda_{23}}&\frac{1}{C_3^2} Y_{\lambda_{33}}&...\\
...&...&...&...\\
\end{pmatrix}.
\end{aligned}
\end{equation}

To obtain the matrix elements of $[\textbf{I}-i \bar{\textbf{R}}]^{-1}$, which are necessary for evaluating the $S$-matrix in MQDT, we employ the Schur block inversion method. If a matrix $\textbf{M}$ is decomposed into four subblocks
\begin{equation}
\begin{aligned}
\textbf{M}=
 \begin{pmatrix} 
\textbf{A}&\textbf{B}\\
\textbf{C}&\textbf{D}
\end{pmatrix}
\end{aligned}
\end{equation}
then its inverse can be written as
\begin{equation}
\begin{aligned}
\textbf{M}^{-1}=
 \begin{pmatrix} 
(\textbf{M}/\textbf{D})^{-1}&-(\textbf{M}/\textbf{D})^{-1}\textbf{B}\textbf{D}^{-1}\\
-\textbf{D}^{-1} \textbf{C}(\textbf{M}/\textbf{D})^{-1}&\textbf{D}^{-1}+\textbf{D}^{-1}\textbf{C}(\textbf{M}/\textbf{D})^{-1}\textbf{B} \textbf{D}^{-1}
\end{pmatrix}\\
\end{aligned}
\end{equation}
where
\begin{equation}
\begin{aligned}
\textbf{M}/\textbf{D}=\textbf{A}-\textbf{B} \textbf{D}^{-1} \textbf{C}.
\end{aligned}
\end{equation} 
is the Schur complement of the block $\mathbf{D}$.

\subsection{The energy dependence of the multichannel $S$-matrix in MQDT}

{We are interested in finding the inverse of the matrix}
\begin{equation}
\begin{aligned}
{
\textbf{I}-i\bar{\textbf{R}} = \begin{pmatrix} 
1-i\frac{1}{C_1^2}Y_{\lambda_{11}}&-i\frac{1}{C_1 C_2}Y_{\lambda_{12}} &-i\frac{1}{C_1 C_3} Y_{\lambda_{13}} &...\\
-i\frac{1}{C_1 C_2} Y_{\lambda_{12}}&1-i\frac{1}{C_2^2} Y_{\lambda_{22}}&-i\frac{1}{C_2 C_3} Y_{\lambda_{23}}&...\\
-i\frac{1}{C_1 C_3} Y_{\lambda_{13}}&-i\frac{1}{C_2 C_3} Y_{\lambda_{23}}&1-i\frac{1}{C_3^2} Y_{\lambda_{33}}&...\\
...&...&...&...\\
\end{pmatrix}}
\end{aligned}
\end{equation}

To apply  Schur's  method, we must first define the subblocks $\textbf{A}$, $\textbf{B}$, $\textbf{C}$, and $\textbf{D}$.  It is advantageous to  separate the energy-dependent and energy-independent parts of $\textbf{I}-i\bar{\textbf{R}}$. The only energy-dependent term is $C_1$, which  only occurs in the first row and first column of the matrix. Thus, we choose to define the subblocks as
\begin{equation}
\begin{aligned}
{
\textbf{A}=1-i \frac{1}{C_1^2}Y_{\lambda_{11}}}, \,\,\,
{
\textbf{B}=  
\begin{pmatrix} 
-i\frac{1}{C_1 C_2}Y_{\lambda_{12}} &-i\frac{1}{C_1 C_3} Y_{\lambda_{13}} &-i\frac{1}{C_1 C_4} Y_{\lambda_{14}}&...
\end{pmatrix}},\\
\textbf{C}=  
 \begin{pmatrix} 
-i\frac{1}{C_1 C_2} Y_{\lambda_{12}}\\
-i\frac{1}{C_1 C_3} Y_{\lambda_{13}}\\
-i\frac{1}{C_1 C_4} Y_{\lambda_{14}}\\
...\\
\end{pmatrix}, \qquad
\textbf{D}=  
 \begin{pmatrix} 
1-i\frac{1}{C_2^2} Y_{\lambda_{22}}
&
-i\frac{1}{C_2 C_3} Y_{\lambda_{23}}&-i\frac{1}{C_2 C_4} Y_{\lambda_{24}}&...\\
-i\frac{1}{C_2 C_3}  Y_{\lambda_{23}}&
1-i\frac{1}{C_3^2} Y_{\lambda_{33}}&-i\frac{1}{C_3 C_4} Y_{\lambda_{34}}&...\\
-i\frac{1}{C_2 C_4}  Y_{\lambda_{24}}&
-i\frac{1}{C_3 C_4} Y_{\lambda_{34}}&1-i\frac{1}{ C_4^2} Y_{\lambda_{44}}&...\\
...&...&...&...\\
\end{pmatrix}.
\end{aligned}
\end{equation}

Define the inverse of the energy-independent subblock $\textbf{D}$
\begin{equation}
\begin{aligned}
 \textbf{D}^{-1}=\begin{pmatrix} 
\tilde{D}_{11}&\tilde{D}_{12}&\tilde{D}_{13}&...\\
\tilde{D}_{21}&\tilde{D}_{22}&\tilde{D}_{23}&...\\
\tilde{D}_{31}&\tilde{D}_{32}&\tilde{D}_{33}&...\\
...&...&...&...\\
\end{pmatrix} 
\end{aligned}
\end{equation}
{From now on the term $\tilde{D}$ represents the inverse  of the $\textbf{D}$ matrix.}


 Applying the block inversion method in combination with these equations, the elastic $S$-matrix element can be written as
 \begin{equation}
\begin{aligned}\label{S11}
 S_{11}=e^{2 i \textcolor{magenta}{\xi_1}} \frac{1+ \frac{a_1}{\textcolor{magenta}{C_1^2}} }{1+\frac{1}{\textcolor{magenta}{C_1^2}} M_N},  
\end{aligned}
\end{equation}
where 
\begin{equation}
\begin{aligned}\label{consta}
  a_1 = i Y_{\lambda_{11}} +
i \frac{Y_{\lambda_{12}}}{C_2} (i \tilde{D}_{11} \frac{Y_{\lambda_{12}}}{C_2} +i \tilde{D}_{12} \frac{Y_{\lambda_{13}}}{C_3}+i \tilde{D}_{13} \frac{Y_{\lambda_{14}}}{C_4}+...)\\
+i\frac{Y_{\lambda_{13}}}{C_3} (i \tilde{D}_{12} \frac{Y_{\lambda_{12}}}{C_2}+i \tilde{D}_{22} \frac{Y_{\lambda_{13}}}{C_3}+i \tilde{D}_{32} \frac{Y_{\lambda_{14}}}{C_4}+...)\\
+ i\frac{Y_{\lambda_{14}}}{C_4} (i\tilde{D}_{13} \frac{Y_{\lambda_{12}}}{C_2}+i\tilde{D}_{23} \frac{Y_{\lambda_{13}}}{C_3}+i \tilde{D}_{33} \frac{Y_{\lambda_{14}}}{C_4} \bigg)\\
\\
M_N=-iY_{\lambda_{11}}+
\frac{1}{C_2}Y_{\lambda_{12}}
(\frac{\tilde{D}_{11}}{ C_2} Y_{\lambda_{12}} +\frac{\tilde{D}_{12}}{ C_3} Y_{\lambda_{13}} +\frac{\tilde{D}_{13}}{ C_4} Y_{\lambda_{14}}+... )\\
+\frac{1}{ C_3} Y_{\lambda_{13}}
(\frac{\tilde{D}_{21}}{ C_2} Y_{\lambda_{12}} +\frac{\tilde{D}_{22}}{ C_3} Y_{\lambda_{13}}+\frac{\tilde{D}_{23}}{ C_4} Y_{\lambda_{14}}+...)\\
+\frac{1}{ C_4} Y_{\lambda_{14}}
(\frac{\tilde{D}_{31}}{ C_2} Y_{\lambda_{12}} +\frac{\tilde{D}_{32}}{ C_3} Y_{\lambda_{13}}+\frac{\tilde{D}_{33}}{ C_4} Y_{\lambda_{14}}+...)
 \end{aligned}
\end{equation}
are energy-independent complex constants 

In a similar way, the first row of the $S$-matrix can be written as
\begin{equation}
\begin{aligned}
S_{1n}=e^{i\textcolor{magenta}{\xi_1}} e^{i\xi_n} \bigg[(1+ i \frac{Y_{\lambda_{11}}}{\textcolor{magenta}{C_1^2}})\frac{1}{\textcolor{magenta}{C_1}} \frac{1}{1+\frac{1}{\textcolor{magenta}{C_1^2}} M_N} h_1
+ \bigg(h_1+\frac{1}{\textcolor{magenta}{C_1^2}} \frac{1}{1+\frac{1}{\textcolor{magenta}{C_1^2} } M_N}  h_2 \bigg)\bigg]
 \end{aligned}
\end{equation}
where the energy-independent complex  constants $h_1$ and $h_2$ are given by
\begin{equation}
\begin{aligned}
  h_1=(i\tilde{D}_{1(n-1)} \frac{Y_{\lambda_{12}}}{C_2}+i\tilde{D}_{2(n-1)} \frac{Y_{\lambda_{13}}}{C_3}+i \tilde{D}_{3(n-1)} \frac{Y_{\lambda_{14}}}{C_4}+...)\\
  \\
  h_2=i\frac{Y_{\lambda_{12}}}{C_2}(i\tilde{D}_{1(n-1)} \frac{Y_{\lambda_{12}}}{C_2}+i\tilde{D}_{2(n-1)} \frac{Y_{\lambda_{13}}}{C_3}+i \tilde{D}_{3(n-1)} \frac{Y_{\lambda_{14}}}{C_4}+...) \\(i \tilde{D}_{11} \frac{Y_{\lambda_{12}}}{C_2} +i \tilde{D}_{12} \frac{Y_{\lambda_{13}}}{C_3}+i \tilde{D}_{13} \frac{Y_{\lambda_{14}}}{C_4}+...)\\
+i\frac{Y_{\lambda_{13}}}{C_3}(i\tilde{D}_{1(n-1)} \frac{Y_{\lambda_{12}}}{C_2}+i\tilde{D}_{2(n-1)} \frac{Y_{\lambda_{13}}}{C_3}+i \tilde{D}_{3(n-1)} \frac{Y_{\lambda_{14}}}{C_4}+...)\\(i \tilde{D}_{12} \frac{Y_{\lambda_{12}}}{C_2}+i \tilde{D}_{22} \frac{Y_{\lambda_{13}}}{C_3}+i \tilde{D}_{32} \frac{Y_{\lambda_{14}}}{C_4}+...) \\
+ i\frac{Y_{\lambda_{14}}}{C_4}(i\tilde{D}_{1(n-1)} \frac{Y_{\lambda_{12}}}{C_2}+i\tilde{D}_{2(n-1)} \frac{Y_{\lambda_{13}}}{C_3}+i \tilde{D}_{3(n-1)} \frac{Y_{\lambda_{14}}}{C_4}+...)\\(i\tilde{D}_{13} \frac{Y_{\lambda_{12}}}{C_2}+i\tilde{D}_{23} \frac{Y_{\lambda_{13}}}{C_3}+i \tilde{D}_{33} \frac{Y_{\lambda_{14}}}{C_4}+...)+...
 \end{aligned}
\end{equation}

The first column of the $S$-matrix can be written as
\begin{equation}
\begin{aligned}
S_{n1}=e^{i\textcolor{magenta}{\xi_1}} e^{i\xi_n} \frac{1}{\textcolor{magenta}{C_1}}\frac{1}{1+\frac{1}{\textcolor{magenta}{C_1^2}} M_N} b_n
 \end{aligned}
\end{equation}
where
\begin{equation}
\begin{aligned}\label{constb}
  b_n= i \frac{Y_{\lambda_{1n}}}{C_n} +(i\tilde{D}_{1(n-1)} \frac{Y_{\lambda_{12}}}{C_2}+i\tilde{D}_{2(n-1)} \frac{Y_{\lambda_{13}}}{C_3}+i \tilde{D}_{3(n-1)} \frac{Y_{\lambda_{14}}}{C_4}+...)\\
  i \frac{Y_{\lambda_{2n}}}{C_2 C_n}  (i \tilde{D}_{11} \frac{Y_{\lambda_{12}}}{C_2} +i \tilde{D}_{12} \frac{Y_{\lambda_{13}}}{C_3}+i \tilde{D}_{13} \frac{Y_{\lambda_{14}}}{C_4}+...)\\
  +  i \frac{Y_{\lambda_{3n}}}{C_3 C_n} (i \tilde{D}_{12} \frac{Y_{\lambda_{12}}}{C_2}+i \tilde{D}_{22} \frac{Y_{\lambda_{13}}}{C_3}+i \tilde{D}_{32} \frac{Y_{\lambda_{14}}}{C_4}+...)\\
  +i\frac{Y_{\lambda_{4n}}}{C_4 C_n} (i\tilde{D}_{13} \frac{Y_{\lambda_{12}}}{C_2}+i\tilde{D}_{23} \frac{Y_{\lambda_{13}}}{C_3}+i \tilde{D}_{33} \frac{Y_{\lambda_{14}}}{C_4}+...)+...\bigg)
 \end{aligned}
\end{equation}

Finally, the rest of the $S$-matrix takes the form 
\begin{equation}
\begin{aligned}
S_{mn}=e^{i\xi_m} e^{i\xi_n}  \bigg( \frac{1}{\textcolor{magenta}{C_1^2}} \frac{1}{1+\frac{1}{\textcolor{magenta}{C_1^2}} M_N} g_1+  g_2\bigg)
\end{aligned}
\end{equation}
where $n\neq1$, $m\neq1$, and
\begin{equation}
\begin{aligned}
  g_1=\bigg(i\frac{Y_{\lambda_{1m}}}{C_m}+(i \tilde{D}_{1(m-1)} \frac{Y_{\lambda_{12}}}{C_2} +i \tilde{D}_{2(m-1)} \frac{Y_{\lambda_{13}}}{C_3}+i \tilde{D}_{3(m-1)} \frac{Y_{\lambda_{14}}}{C_4}+...)\bigg)\\
   (i \tilde{D}_{1(n-1)} \frac{Y_{\lambda_{12}}}{C_2}+i \tilde{D}_{2(n-1)} \frac{Y_{\lambda_{13}}}{C_3}+i \tilde{D}_{3(n-1)} \frac{Y_{\lambda_{14}}}{C_4}+...)\\
 + (i \tilde{D}_{1(n-1)} \frac{Y_{\lambda_{12}}}{C_2}+i \tilde{D}_{2(n-1)} \frac{Y_{\lambda_{13}}}{C_3}+i \tilde{D}_{3(n-1)} \frac{Y_{\lambda_{14}}}{C_4}+...)  \\
 \bigg(i\frac{Y_{\lambda_{2m}}}{C_2 c_m} (i \tilde{D}_{11} \frac{Y_{\lambda_{12}}}{C_2} +i \tilde{D}_{12} \frac{Y_{\lambda_{13}}}{C_3}+i \tilde{D}_{13} \frac{Y_{\lambda_{14}}}{C_4}+...)\\
 +i\frac{Y_{\lambda_{3m}}}{C_3 c_m}  (i \tilde{D}_{12} \frac{Y_{\lambda_{12}}}{C_2} +i \tilde{D}_{22} \frac{Y_{\lambda_{13}}}{C_3}+i \tilde{D}_{23} \frac{Y_{\lambda_{14}}}{C_4}+...)\\
 +i\frac{Y_{\lambda_{4m}}}{C_4 c_m}  (i \tilde{D}_{13} \frac{Y_{\lambda_{12}}}{C_2} +i \tilde{D}_{23} \frac{Y_{\lambda_{13}}}{C_3}+i \tilde{D}_{33} \frac{Y_{\lambda_{14}}}{C_4}+...)+...)\bigg)\\
  \\
  g_2=
  (i \tilde{D}_{1(n-1)} \frac{Y_{\lambda_{m2}}}{C_2C_m}+i \tilde{D}_{2(n-1)} \frac{Y_{\lambda_{m3}}}{C_3 C_m}+i \tilde{D}_{3(n-1)} \frac{Y_{\lambda_{m4}}}{C_4 C_m}+...)  
 \end{aligned}
\end{equation}
We simplify these equations by defining a new energy-dependent variable 
$f(\textcolor{magenta}{C_1^2})= \frac{1}{1+\frac{1}{\textcolor{magenta}{C_1^2}}M}$,
in terms of which the scattering matrix elements can be written as
\begin{equation}
\begin{aligned}
S_{11}=e^{2 i\textcolor{magenta}{\textcolor{magenta}{\xi_1}}} f(\textcolor{magenta}{C_1^2})\bigg( 1+ \frac{a_1}{\textcolor{magenta}{C_1^2}} \bigg),
\\
S_{n1}=e^{i\textcolor{magenta}{\xi_1}} e^{i\xi_n}
 f(\textcolor{magenta}{C_1^2}) \frac{b_1}{\textcolor{magenta}{C_1} },  \\
 \\
S_{1n}=e^{i\textcolor{magenta}{\xi_1}} e^{i\xi_n} \bigg((1+ i \frac{Y_{\lambda_{11}}}{\textcolor{magenta}{C_1^2}})\frac{1}{\textcolor{magenta}{C_1}} f(\textcolor{magenta}{C_1^2})h_1
+ \bigg(h_1+\frac{1}{\textcolor{magenta}{C_1^2}} f(\textcolor{magenta}{C_1^2} )h_2 \bigg)\bigg),\\
\\
S_{mn}=e^{i\xi_m} e^{i\xi_n}  \bigg( \frac{1}{\textcolor{magenta}{C_1^2}}f(\textcolor{magenta}{C_1^2}) g_1+  g_2\bigg).
\end{aligned}
\end{equation}

To find $\textbf{Q}$, we also need the  derivative of the $S$-matrix with respect to energy.  From the expressions derived above, we readily obtain the energy derivative of the elastic $S$-matrix element 
 \begin{equation}\label{s11der}
\begin{aligned} 
\frac{d S_{11}}{d\textcolor{magenta}{E}}=2 i \textcolor{magenta}{\xi_1}^\prime e^{2 i \textcolor{magenta}{\xi_1}} \frac{1+ \frac{a_1}{\textcolor{magenta}{C_1^2}}}{1+\frac{1}{\textcolor{magenta}{C_1^2}} M_N}  +e^{2 i \textcolor{magenta}{\xi_1}} \frac{ 2 \textcolor{magenta}{C_1}  \textcolor{magenta}{C_1^\prime} (M_N-a_1) }{(\textcolor{magenta}{C_1^2} + M_N)^2},
\end{aligned}
\end{equation}
as well as the energy derivative of the first column of the $S$-matrix
\begin{equation}
\begin{aligned}
\frac{d S_{n1}}{d\textcolor{magenta}{E}}
=i \textcolor{magenta}{\xi_1^\prime} e^{i \textcolor{magenta}{\xi_1}} e^{i\xi_n}   \frac{1}{\textcolor{magenta}{C_1}}\frac{1}{1+\frac{1}{\textcolor{magenta}{C_1^2}} M_N} b_n+ e^{i\textcolor{magenta}{\xi_1}}e^{i\xi_n}  \frac{\textcolor{magenta}{C_1}^\prime (M_N-\textcolor{magenta}{C_1^2})}{(\textcolor{magenta}{C_1^2}+ M_N)^2} b_n
 \end{aligned}
\end{equation}
These expressions are sufficient to obtain the average time delay in the incident collision channel ($Q_{11}$), which is of primary interest in this work.


\subsection{Energy dependence of $\textbf{S}$  and $d\textbf{S}/dE$ in the s-wave regime}

To obtain the explicit energy dependence of $\textbf{S}$  and $d\textbf{S}/dE$, we need to specify the quantum defect parameters $C_1$ and $\xi_1$ as a function of energy. In the s-wave regime ($k\to0$ and $l=0$) these parameters are  \cite{Idziaszek:10}
\begin{equation}
\begin{aligned}\label{swave}
C^{-2}=k\bar{a}[1+(s-1)^2],\\
\tan\xi=-ka,\\
\tan\lambda=1-s=1-\frac{a}{\bar{a}},
 \end{aligned}
\end{equation}
where $k=\sqrt{2\mu E}$, $s={a}/{\bar{a}}$ is the dimensionless scatting length,  $a$ is the scattering length, $\bar{a}=4 \pi C_6/\Gamma(\frac{1}{4})^2$  is the mean scattering length, and $C_6$ is the long-range dispersion coefficient \cite{Idziaszek:10}.
 Using these definitions $S_{11}$  can be written as
  \begin{equation}
\begin{aligned} 
S_{11}
=e^{2 i \atan(-a \textcolor{magenta}{k})} \frac{1+ \textcolor{magenta}{k}\bar{a} [1+(s-1)^2] a_1}{1+\textcolor{magenta}{k}\bar{a} [1+(s-1)^2] M_N}  
\end{aligned}
\end{equation}
 In the low-energy regime $ka\to 0$  and $S_{11}$ becomes
  \begin{equation}
\begin{aligned} 
\lim_{k\to0} S_{11}=\lim_{k\to0} e^{2 i \atan(-a \textcolor{magenta}{k})} \frac{1+ \textcolor{magenta}{k}\bar{a} [1+(s-1)^2] a_1}{1+\textcolor{magenta}{k}\bar{a} [1+(s-1)^2] M_N}  =e^{-2 i  a \textcolor{magenta}{k}} 
\end{aligned}
\end{equation}

The energy derivative of $S_{11}$ in the s-wave regime is
 \begin{equation}
\begin{aligned} 
\frac{d S_{11}}{d\textcolor{magenta}{E}}
=e^{2  i \atan(-a \textcolor{magenta}{k} )}\bigg(\frac{ -{2 ia \mu} }{(1+ a^2  \textcolor{magenta}{k^2})\textcolor{magenta}{k}}  \frac{1+ \textcolor{magenta}{k}\bar{a} [1+(s-1)^2]a_1 }{1+\textcolor{magenta}{k}\bar{a} [1+(s-1)^2] M_N}  \\
  +\frac{  {\mu}(a_1-M_N)\bar{a} [1+(s-1)^2]}{\textcolor{magenta}{k}(1+\textcolor{magenta}{k}\bar{a} [1+(s-1)^2] M_N)^2}\bigg)
\end{aligned}
\end{equation}

In the limit $ka \to 0$ the energy derivative of $S_{11}$ becomes
 \begin{equation}
\begin{aligned} 
\lim_{k\to0} \frac{d S_{11}}{d\textcolor{magenta}{E}}=\frac{e^{-2  i a \textcolor{magenta}{k} }}{\textcolor{magenta}{k}}\bigg( -\frac{2 a i \mu}{\hbar^2}  
-   \frac{\mu}{\hbar^2}
  (M_N-a_1) \bigg)
\end{aligned}
\end{equation}

Similarly, for the first column of the $S$-matrix  in the s-wave regime, we have
  \begin{equation}
\begin{aligned} 
S_{n1}=
e^{i\atan(-a \textcolor{magenta}{k})} e^{i\xi_n}   \sqrt{\textcolor{magenta}{k}\bar{a} [1+(s-1)^2]}\frac{1}{1+\textcolor{magenta}{k}\bar{a} [1+(s-1)^2] M_N} b_n
\end{aligned}
\end{equation}
As $ka \to 0$ the matrix elements $S_{n1}$ become
  \begin{equation}
\begin{aligned} 
\lim_{k\to0}  S_{n1}=e^{- i a \textcolor{magenta}{k}} e^{i\xi_n}   
\sqrt{\textcolor{magenta}{k}\bar{a} [1+(s-1)^2]} b_n
\end{aligned}
\end{equation}
 
Thus, the energy derivative of $S_{n1}$ in the s-wave regime is (for $n\ne 1$)
\begin{equation}
\begin{aligned}
\frac{d S_{n1}}{d\textcolor{magenta}{E}}=e^{i \atan(-a \textcolor{magenta}{k} )} e^{i\xi_n}   b_n\bigg(  \frac{ -{i a \mu} }{(1+ a^2  \textcolor{magenta}{k^2})\textcolor{magenta}{k}}
  \frac{\sqrt{\textcolor{magenta}{k}\bar{a} [1+(s-1)^2]}}{1+ \textcolor{magenta}{k}\bar{a} [1+(s-1)^2] M_N} \\
+   \frac{{\mu}\frac{-\textcolor{magenta}{k}\bar{a} [1+(s-1)^2]}{ 2 \textcolor{magenta}{k^{5/2}}\sqrt{\bar{a} [1+(s-1)^2]}}
  (\textcolor{magenta}{k}\bar{a} [1+(s-1)^2] M_N-1)}{(1+\textcolor{magenta}{k}\bar{a} [1+(s-1)^2] M_N)^2} \bigg).
 \end{aligned}
\end{equation}

In the limit $ka\to 0$ we have
\begin{equation}
\begin{aligned}
\lim_{k\to0} \frac{d S_{n1}}{d{E}}
=e^{- i a \textcolor{magenta}{k} } e^{i\xi_n}   b_n \sqrt{\bar{a} [1+(s-1)^2]}\bigg(  \frac{ -{i a \mu} }{\textcolor{magenta}{k^{1/2}}} 
+  {\mu}\frac{1}{ 2 \textcolor{magenta}{k^{3/2}} }
 \bigg)
 \end{aligned}
\end{equation}

The expressions derived in this section are used below to calculate the average dime delay in the incident collision channel $Q_{11}$.

\subsection{Energy dependence of $\mathbf{S}$  and $d\mathbf{S}/dE$ in the p-wave regime}

It is also interesting to explore the $S$-matrix and its energy derivative in the p-wave regime. 
Similarly to the s-wave case, we start by expressing the quantum defect parameters as a function of energy \cite{Idziaszek:10} 
\begin{equation}
\begin{aligned}
C^{-2}=2.128 \textcolor{magenta}{k^3}\bar{a}^3 \frac{1+(s-1)^2}{(s-2)^2},\\
\tan\xi=2.128  \textcolor{magenta}{k^3}  a^3 \frac{s-1}{s-2},\\
\tan\lambda=\frac{s}{s-2}.
 \end{aligned}
\end{equation}
In the p-wave regime, $S_{11}$ can be represented as 
\begin{equation}
\begin{aligned}
 S_{11}=e^{2 i \atan(2.128  \textcolor{magenta}{k^3}  a^3 \frac{s-1}{s-2})} \frac{1+ 2.128 \textcolor{magenta}{k^3}\bar{a}^3 \frac{1+(s-1)^2}{(s-2)^2}a_1 }{1+2.128 \textcolor{magenta}{k^3}\bar{a}^3 \frac{1+(s-1)^2}{(s-2)^2} M_N}
  \end{aligned}
\end{equation} 
As $ka$ goes to zero, the  matrix element  $S_{11}$ becomes
\begin{equation}
\begin{aligned}
 \lim_{k\to0} S_{11}=e^{2 i\atan(2.128  \textcolor{magenta}{k^3}  a^3 \frac{s-1}{s-2})}.
  \end{aligned}
\end{equation} 

Taking the energy derivative of $S_{11}$ we find
 \begin{equation}
\begin{aligned} 
\frac{d S_{11}}{d\textcolor{magenta}{E}}=e^{2 i \atan(2.128  \textcolor{magenta}{k^3}  a^3 \frac{s-1}{s-2})}\bigg(2 i \frac{2.128 \frac{3 \mu}{\hbar^2}   a^3 \frac{s-1}{s-2} \textcolor{magenta}{k} }{1+(2.128   a^3 \frac{s-1}{s-2} )^2  \textcolor{magenta}{k^6}}
\frac{1+ 2.128 \textcolor{magenta}{k^3}\bar{a}^3 \frac{1+(s-1)^2}{(s-2)^2}a_1 }{1+2.128 \textcolor{magenta}{k^3}\bar{a}^3 \frac{1+(s-1)^2}{(s-2)^2} M_N}  \\
+ \frac{ 3 \frac{\mu}{\hbar^2} \textcolor{magenta}{k} 2.128 \bar{a}^3 \frac{1+(s-1)^2}{(s-2)^2}(a_1-M_N)}{(1+2.128 \textcolor{magenta}{k^3}\bar{a}^3 \frac{1+(s-1)^2}{(s-2)^2} M_N)^2} \bigg)
\end{aligned}
\end{equation}

As $ka$ approaches zero, the derivative  becomes
 \begin{equation}
\begin{aligned} 
\lim_{k\to0} \frac{d S_{11}}{d\textcolor{magenta}{E}}=e^{4.256 i  \textcolor{magenta}{k^3}  a^3 \frac{s-1}{s-2}}\bigg( i 2.128 \frac{6 \mu}{\hbar^2}   a^3 \frac{s-1}{s-2} \textcolor{magenta}{k} \\ + \frac{3\mu}{\hbar^2} 
2.128 \textcolor{magenta}{k}\bar{a}^3 \frac{1+(s-1)^2}{(s-2)^2} (a_1- M_N) \bigg)
\end{aligned}
\end{equation}

In the p-wave regime $S_{n1}$ can be represented as 
\begin{equation}
\begin{aligned}
 S_{n1}=e^{i \atan(2.128  \textcolor{magenta}{k^3}  a^3 \frac{s-1}{s-2})} e^{i\xi_n} \sqrt{2.128 \textcolor{magenta}{k^3}\bar{a}^3 \frac{1+(s-1)^2}{(s-2)^2}}\frac{1}{1+2.128 \textcolor{magenta}{k^3}\bar{a}^3 \frac{1+(s-1)^2}{(s-2)^2} M_N} b_n
  \end{aligned}
\end{equation} 
as $ka$ goes to 0, $S_{n1}$ matrix element becomes
\begin{equation}
\begin{aligned}
 \lim_{k\to0}S_{n1}=e^{i \atan(2.128  \textcolor{magenta}{k^3}  a^3 \frac{s-1}{s-2})} e^{i\xi_n} \sqrt{2.128 \textcolor{magenta}{k^3}\bar{a}^3 \frac{1+(s-1)^2}{(s-2)^2}} b_n
  \end{aligned}
\end{equation} 
 
The energy derivative of $S_{n1}$ ($n\ne 1$) in the p-wave regime takes the form
\begin{equation}
\begin{aligned}
\frac{d S_{n1}}{d\textcolor{magenta}{E}}
=e^{i \atan(2.128  \textcolor{magenta}{k^3}  a^3 \frac{s-1}{s-2})} e^{i\xi_n}   b_n\bigg( i \frac{2.128 {3 \mu}   a^3 \frac{s-1}{s-2} \textcolor{magenta}{k} }{1+(2.128   a^3 \frac{s-1}{s-2} )^2  \textcolor{magenta}{k^6}}
  \frac{\sqrt{2.128 \textcolor{magenta}{k^3}\bar{a}^3 \frac{1+(s-1)^2}{(s-2)^2}}}{1+ 2.128 \textcolor{magenta}{k^3}\bar{a}^3 \frac{1+(s-1)^2}{(s-2)^2}M_N} \\
+   \frac{{\mu}\frac{-3}{ 2 \textcolor{magenta}{k^{1/2}}}
  \sqrt{2.128 \bar{a}^3 \frac{1+(s-1)^2}{(s-2)^2}}(M_N 2.128 \textcolor{magenta}{k^3}\bar{a}^3 \frac{1+(s-1)^2}{(s-2)^2}-1)}{(1+2.128 \textcolor{magenta}{k^3}\bar{a}^3 \frac{1+(s-1)^2}{(s-2)^2} M_N)^2} \bigg).
 \end{aligned}
\end{equation}
As $ka$ approaches zero, the energy derivative  can be written as
\begin{equation}
\begin{aligned}
\lim_{k\to0}  \frac{d S_{n1}}{d\textcolor{magenta}{E}}
=e^{i 2.128  \textcolor{magenta}{k^3}  a^3 \frac{s-1}{s-2}} e^{i\xi_n}   b_n\bigg( i 2.128 {3 \mu}   a^3 \frac{s-1}{s-2} \textcolor{magenta}{k^{5/2}} 
\sqrt{2.128 \bar{a}^3 \frac{1+(s-1)^2}{(s-2)^2}}\\
+{\mu} \frac{3\sqrt{2.128 \bar{a}^3 \frac{1+(s-1)^2}{(s-2)^2}} }{ 2 \textcolor{magenta}{k^{1/2}}}
 \bigg).
 \end{aligned}
\end{equation}

\section{\label{Qmat_MQDT_param}{The Q-matrix in terms of MQDT parameters}}

{Substituting MQDT expressions for the S-matrix elements (\ref{S11}) and their energy derivatives  (\ref{s11der}) in  Eq.~\eqref{Smith}, we finally  obtain the desired result for the average time delay in the incident channel ($i=1$) in the absence of closed channels} 
\begin{equation}
 \begin{aligned}\label{Q}
Q_{11}=2  \textcolor{magenta}{\xi_1}^\prime \frac{|1+ \frac{a_1}{\textcolor{magenta}{C_1^2}}|^2}{|1+\frac{1}{\textcolor{magenta}{C_1^2}} M_N|^2}  + i  \frac{ 2 \textcolor{magenta}{C_1}  \textcolor{magenta}{C_1^\prime}(2M_N^*) (1- \frac{M_N}{\textcolor{magenta}{C_1^2}})}{(1+\frac{1}{\textcolor{magenta}{C_1^2}} M_N^*) |\textcolor{magenta}{C_1^2}+ M_N|^2} \\
+ |b_n|^2  (
 \textcolor{magenta}{\xi_1^\prime}  \frac{1}{\textcolor{magenta}{C_1}^2}
\frac{1}{|1+\frac{1}{\textcolor{magenta}{C_1^2}} M_N|^2} +i  \frac{\textcolor{magenta}{C_1} \textcolor{magenta}{C_1}^\prime(\frac{1}{\textcolor{magenta}{C_1^2}}M_N^*-1)}{|(\textcolor{magenta}{C_1^2}+ M_N)|^2 (1+M_N^*\frac{1}{\textcolor{magenta}{C_1^2}}) } ).\\
 \end{aligned}
\end{equation}
The first two terms in equation \eqref{Q} come from $i S_{11}^\dagger \frac{d S_{11}}{dE}$ and hence represent elastic contributions. The second two terms come from $i S_{n1}^\dagger \frac{d S_{n1}}{dE}$, and thereby represent contributions from inelastic scattering.

Because the $S$-matrix is unitary, $Q_{11}$ must be real \cite{Smith:60}),
so we need to show that the imaginary terms in Eq.~\eqref{Q} cancel. In order for this to happen,   the following equations must be true: 
  \begin{equation}
\begin{aligned} 
 i \frac{ 2 \textcolor{magenta}{C_1}  \textcolor{magenta}{C_1^\prime}(2M_N^*) (1- \frac{M_N}{\textcolor{magenta}{C_1^2}})}{(1+\frac{1}{\textcolor{magenta}{C_1^2}} M_N^*) |\textcolor{magenta}{C_1^2}+ M_N|^2} 
=-
 i (\sum_n |b_n|^2  )
  \frac{\textcolor{magenta}{C_1} \textcolor{magenta}{C_1}^\prime(\frac{1}{\textcolor{magenta}{C_1^2}}M_N^*-1)}{|(\textcolor{magenta}{C_1^2}+ M_N)|^2 (1+M_N^*\frac{1}{\textcolor{magenta}{C_1^2}}) } \\
   i \frac{ 2 \textcolor{magenta}{C_1}  \textcolor{magenta}{C_1^\prime}(2M_N^*) (1- \frac{M_N}{\textcolor{magenta}{C_1^2}}) (1+\frac{1}{\textcolor{magenta}{C_1^2}} M_N)}{|(1+\frac{1}{\textcolor{magenta}{C_1^2}} M_N^*)|^2 |\textcolor{magenta}{C_1^2}+ M_N|^2} 
=
 i (\sum_n |b_n|^2  )
  \frac{\textcolor{magenta}{C_1} \textcolor{magenta}{C_1}^\prime(1-\frac{1}{\textcolor{magenta}{C_1^2}}M_N^*)(1+\frac{1}{\textcolor{magenta}{C_1^2}} M_N)}{|(\textcolor{magenta}{C_1^2}+ M_N)|^2 |(1+M_N^*\frac{1}{\textcolor{magenta}{C_1^2}})|^2 } 
 \end{aligned}
\end{equation}
Since the denominators of the fractions are real and equal to each other,  we can cancel them:
  \begin{equation}
\begin{aligned} 
   i  4 \textcolor{magenta}{C_1}  \textcolor{magenta}{C_1^\prime} M_N^* (1- \frac{M_N}{\textcolor{magenta}{C_1^2}}) (1+\frac{1}{\textcolor{magenta}{C_1^2}} M_N)
=
 i (\sum_n |b_n|^2  )
\textcolor{magenta}{C_1} \textcolor{magenta}{C_1}^\prime(1-\frac{1}{\textcolor{magenta}{C_1^2}}M_N^*)(1+\frac{1}{\textcolor{magenta}{C_1^2}} M_N)
 \end{aligned}
\end{equation}
Additionally, we divide both sides of the equation by $C_1$ and $C_1^\prime$ 
 \begin{equation}
\begin{aligned} 
   i  4 M_N^* (1- \frac{M_N}{\textcolor{magenta}{C_1^2}}) (1+\frac{1}{\textcolor{magenta}{C_1^2}} M_N)
=
 i (\sum_n |b_n|^2  )
(1-\frac{1}{\textcolor{magenta}{C_1^2}}M_N^*)(1+\frac{1}{\textcolor{magenta}{C_1^2}} M_N)
 \end{aligned}
\end{equation}
and multiply to obtain
 \begin{equation}
\begin{aligned} 
   i  4 M_N^* (1-\frac{1}{\textcolor{magenta}{C_1^4}}M_N^{2}) =
 i (\sum_n |b_n|^2  ) (1+ \frac{M_N}{\textcolor{magenta}{C_1^2}}-\frac{1}{\textcolor{magenta}{C_1^2}} M_N^*-\frac{1}{\textcolor{magenta}{C_1^4}} M_N^* M_N)
 \end{aligned}
\end{equation}
Now we express $M_N$ in terms of its real and imaginary parts.
 \begin{equation}
\begin{aligned} 
   i  4 (M_R-i M_I) (1-\frac{1}{\textcolor{magenta}{C_1^4}}(M_R+i M_I)^2)\\
= i (\sum_n |b_n|^2  )
(1+ \frac{(M_R+i M_I)}{\textcolor{magenta}{C_1^2}}-\frac{(M_R-i M_I)}{\textcolor{magenta}{C_1^2}} -\frac{1}{\textcolor{magenta}{C_1^4}} (M_R-i M_I) (M_R+i M_I))
 \end{aligned}
\end{equation}
Expanding the square, we obtain 
 \begin{equation}
\begin{aligned} 
    4 (iM_R+ M_I- \frac{1}{\textcolor{magenta}{C_1^4}} (i M_R^3- M_R^2 M_I - M_I^3+i M_R M_I^2))\\
=
  (\sum_n |b_n|^2  )
(i-2 \frac{ M_I}{\textcolor{magenta}{C_1^2}}- i\frac{1}{\textcolor{magenta}{C_1^4}} (M_R^2+ M_I^2))
 \end{aligned}
\end{equation}
We  require the imaginary parts to cancel:
 \begin{equation}
\begin{aligned} 
    4 (iM_R- \frac{1}{\textcolor{magenta}{C_1^4}} (i M_R^3+i M_R M_I^2))\\
=
  (\sum_n |b_n|^2  )
(i- i\frac{1}{\textcolor{magenta}{C_1^4}} (M_R^2+ M_I^2))
 \end{aligned}
\end{equation}
Since $C_1$ is energy-dependent, both sides of the equation with the same power of $C_1$ must be equal, so we obtain the following equations
 \begin{equation}
\begin{aligned} 
     4 iM_R=i   \sum_n |b_n|^2,\\
     -4i  ( M_R^3+  M_R M_I^2)
=-i\left(\sum_n |b_n|^2  \right) (M_R^2+M_I^2).
 \end{aligned}
\end{equation}
The equations can be simplified  to get:
 \begin{equation}
\begin{aligned} 
     4 M_R=   \sum_n |b_n|^2,  \\
     4 ( M_R^3+  M_R M_I^2)
=\left(\sum_n |b_n|^2  \right) (M_R^2+ M_I^2).
 \end{aligned}
\end{equation}
Both of these equations give $ 4 M_R=   \sum_n |b_n|^2$, which can be verified by comparing $M_R$ and $\sum_n |b_n|^2$. Thus our equation can be rewritten as 
\begin{equation}
\begin{aligned}\label{q1111}
Q_{11}=  2   \textcolor{magenta}{\xi_1^\prime} \frac{|1+ \frac{a_1}{\textcolor{magenta}{C_1^2}}|^2}{|1+\frac{1}{\textcolor{magenta}{C_1^2}} M_N|^2}  +  \frac{ 4 \textcolor{magenta}{C_1}  \textcolor{magenta}{C_1^\prime}(M_I+\frac{1}{\textcolor{magenta}{C_1^4}}(M_R^2 M_I+M_I^3))}{|1+\frac{1}{\textcolor{magenta}{C_1^2}} M_N^*|^2 |\textcolor{magenta}{C_1^2}+ M_N|^2} \\
+  (\hbar
 \textcolor{magenta}{\xi_1^\prime}  \frac{1}{\textcolor{magenta}{C_1}^2}
\frac{4 M_R}{|1+\frac{1}{\textcolor{magenta}{C_1^2}} M_N|^2} + \frac{8 \hbar \textcolor{magenta}{C_1^\prime} M_R M_I  }{\textcolor{magenta}{C_1}  |\textcolor{magenta}{C_1^2}+ M_N|^2 |1+  M_N^*\frac{1}{\textcolor{magenta}{C_1^2}}|^2 } )\\
 \end{aligned}
\end{equation}
which gives us $Q_{11}$ in terms of energy-dependent MQDT parameters.


\subsection{Average  time delay in the s-wave regime}

To find the energy dependence of the average time delay in the s-wave regime, we first rearrange Eq.~\eqref{q1111} 
\begin{equation}\label{q11expanded}
\begin{aligned}
Q_{11}=\frac{1}{|1+\frac{1}{\textcolor{magenta}{C_1^2}} M_N|^2} \bigg(
2  \textcolor{magenta}{\xi_1^\prime} (|1+ \frac{1}{\textcolor{magenta}{C_1^2}}a_1|^2) +  \frac{ 4 \textcolor{magenta}{C_1}  \textcolor{magenta}{C_1^\prime}(M_I+\frac{1}{\textcolor{magenta}{C_1^4}}(M_R^2 M_I+M_I^3))}{(|\textcolor{magenta}{C_1^2}+ M_N|^2)} \\
+  (
 \textcolor{magenta}{\xi_1^\prime}  \frac{1}{\textcolor{magenta}{C_1^2}}
4 M_R + \frac{8 \textcolor{magenta}{C_1^\prime} M_R M_I  }{\textcolor{magenta}{C_1}  (|\textcolor{magenta}{C_1^2}+ M_N|^2)  } )\bigg)
\end{aligned}
\end{equation}
and define the following auxiliary quantities
\begin{equation}
\begin{aligned}
  M_N=M_R+i M_I,\\
  a_1=-M_R+i M_I,\\
  M_I=Y_{\lambda_{11}},\\
  M_R=\Big(
\frac{1}{C_2}Y_{\lambda_{12}}
(\frac{\tilde{D}_{11}}{ C_2} Y_{\lambda_{12}} +\frac{\tilde{D}_{12}}{ C_3} Y_{\lambda_{13}} +\frac{\tilde{D}_{13}}{ C_4} Y_{\lambda_{14}}+... )\\
+\frac{1}{ C_3} Y_{\lambda_{13}}
(\frac{\tilde{D}_{21}}{ C_2} Y_{\lambda_{12}} +\frac{\tilde{D}_{22}}{ C_3} Y_{\lambda_{13}}+\frac{\tilde{D}_{23}}{ C_4} Y_{\lambda_{14}}+...)\\
+\frac{1}{ C_4} Y_{\lambda_{14}}
(\frac{\tilde{D}_{31}}{ C_2} Y_{\lambda_{12}} +\frac{\tilde{D}_{32}}{ C_3} Y_{\lambda_{13}}+\frac{\tilde{D}_{33}}{ C_4} Y_{\lambda_{14}}+...)\Big).
 \end{aligned}
\end{equation}

In the s-wave threshold regime,   the energy derivatives of $\xi=\atan(-a \textcolor{magenta}{k})$ and $C_1^{-2}=\textcolor{magenta}{k}\bar{a} [1+(s-1)^2]$ are given by 
  \begin{equation}
\begin{aligned} 
\frac{d \textcolor{magenta}{\xi}}{d\textcolor{magenta}{E}}=-\frac{ a { \mu} }{ (1+   a^2 \textcolor{magenta}{k^2})\textcolor{magenta}{k}},\\
\frac{d \textcolor{magenta}{C_1^{-2}}}{d\textcolor{magenta}{E}}={\mu} \textcolor{magenta}{k^{-1}}\bar{a} [1+(s-1)^2],\\
\frac{d \textcolor{magenta}{C_1}}{d\textcolor{magenta}{E}}={\mu}\frac{-1}{ 2 \textcolor{magenta}{k^{5/2}}\sqrt{\bar{a} [1+(s-1)^2]}}.
\end{aligned}
\end{equation}
{
We substitute $C_1$ into equation \eqref{q11expanded} and obtain
 \begin{equation}
\begin{aligned}
Q_{11}=\frac{1}{1+2M_R \textcolor{magenta}{k }\bar{a} [1+(s-1)^2]+\textcolor{magenta}{k^2 }\bar{a^2} [1+(s-1)^2]^2 |M_N|^2}\\
 \bigg(
2  \textcolor{magenta}{\xi_1^\prime} (1-2M_R \textcolor{magenta}{k }\bar{a} [1+(s-1)^2]+ \textcolor{magenta}{k^2 }\bar{a^2} [1+(s-1)^2]^2 |a_1|^2) \\
+  \frac{ 4   \textcolor{magenta}{C_1^\prime}(M_I+\textcolor{magenta}{k^2 }\bar{a^2} [1+(s-1)^2]^2(M_R^2 M_I+M_I^3))(\textcolor{magenta}{k^{3/2} }\bar{a^{3/2}} [1+(s-1)^2]^{3/2})}{(1+2M_R \textcolor{magenta}{k }\bar{a} [1+(s-1)^2]+ |M_N|^2 \textcolor{magenta}{k^2 }\bar{a^2} [1+(s-1)^2]^2)} \\
+  (
 \textcolor{magenta}{\xi_1^\prime} \textcolor{magenta}{k }\bar{a} [1+(s-1)^2]
4 M_R \\
+ \frac{8 \textcolor{magenta}{C_1^\prime} M_R M_I  (\textcolor{magenta}{k^{5/2} }\bar{a^{5/2}} [1+(s-1)^2]^{5/2}))}{  (1+2M_R \textcolor{magenta}{k }\bar{a} [1+(s-1)^2]+ |M_N|^2 \textcolor{magenta}{k^{2} }\bar{a^{2}} [1+(s-1)^2]^{2})  } )\bigg)
 \end{aligned}
\end{equation}}

{
Next we substitute the expressions for $C_1^\prime$ and $\xi_1^\prime$ to find
 \begin{equation}\label{q11c1c1prime}
\begin{aligned}
Q_{11}=\frac{1}{1+2M_R \textcolor{magenta}{k }\bar{a} [1+(s-1)^2]+\textcolor{magenta}{k^2 }\bar{a^2} [1+(s-1)^2]^2 |M_N|^2} \\
\bigg(
-2 \frac{ a  \mu }{ (1+   a^2 \textcolor{magenta}{k^2})\textcolor{magenta}{k}}(1-2M_R \textcolor{magenta}{k }\bar{a} [1+(s-1)^2]+ \textcolor{magenta}{k^2 }\bar{a^2} [1+(s-1)^2]^2 |a_1|^2) \\
-  \frac{ 2   {\mu}\frac{1}{  \textcolor{magenta}{k^{5/2}}\sqrt{\bar{a} [1+(s-1)^2]}}(M_I+\textcolor{magenta}{k^2 }\bar{a^2} [1+(s-1)^2]^2(M_R^2 M_I+M_I^3))(\textcolor{magenta}{k^{3/2} }\bar{a^{3/2}} [1+(s-1)^2]^{3/2})}{(1+2M_R \textcolor{magenta}{k }\bar{a} [1+(s-1)^2]+ |M_N|^2 \textcolor{magenta}{k^2 }\bar{a^2} [1+(s-1)^2]^2)} \\
-\frac{ a { \mu} }{ (1+   a^2 \textcolor{magenta}{k^2})\textcolor{magenta}{k}} \textcolor{magenta}{k }\bar{a} [1+(s-1)^2]
4 M_R \\
- \frac{4 {\mu}\frac{1}{  \textcolor{magenta}{k^{5/2}}\sqrt{\bar{a} [1+(s-1)^2]}} M_R M_I  (\textcolor{magenta}{k^{5/2} }\bar{a^{5/2}} [1+(s-1)^2]^{5/2}))}{  (1+2M_R \textcolor{magenta}{k }\bar{a} [1+(s-1)^2]+ |M_N|^2 \textcolor{magenta}{k^{2} }\bar{a^{2}} [1+(s-1)^2]^{2})  } )\bigg)\\
\\
=\frac{1}{1+2M_R \textcolor{magenta}{k }\bar{a} [1+(s-1)^2]+\textcolor{magenta}{k^2 }\bar{a^2} [1+(s-1)^2]^2 |M_N|^2} \\
\bigg(
-2 \frac{ a { \mu} }{ (1+   a^2 \textcolor{magenta}{k^2})\textcolor{magenta}{k}}(1-2M_R \textcolor{magenta}{k }\bar{a} [1+(s-1)^2]+ \textcolor{magenta}{k^2 }\bar{a^2} [1+(s-1)^2]^2 |a_1|^2) \\
-  \frac{ 2   {\mu} \bar{a} [1+(s-1)^2] (M_I+\textcolor{magenta}{k^2 }\bar{a^2} [1+(s-1)^2]^2(M_R^2 M_I+M_I^3))}{ \textcolor{magenta}{k} (1+2M_R \textcolor{magenta}{k }\bar{a} [1+(s-1)^2]+ |M_N|^2 \textcolor{magenta}{k^2 }\bar{a^2} [1+(s-1)^2]^2)} \\
-\frac{ a { \mu} }{ 1+   a^2 \textcolor{magenta}{k^2}} \bar{a} [1+(s-1)^2]
4 M_R \\
- \frac{4 {\mu} M_R M_I (\bar{a^{2}} [1+(s-1)^2]^{2}))}{  (1+2M_R \textcolor{magenta}{k }\bar{a} [1+(s-1)^2]+ |M_N|^2 \textcolor{magenta}{k^{2} }\bar{a^{2}} [1+(s-1)^2]^{2})  } )\bigg)
 \end{aligned}
\end{equation}}
{In the limit  $k\to 0$ this equation becomes
\begin{equation}
\begin{aligned}
Q_{11}=
  -\frac{2 a { \mu} }{ \textcolor{magenta}{k}(1+   a^2 \textcolor{magenta}{k^2})} \frac{1-2M_R \textcolor{magenta}{k }\bar{a} [1+(s-1)^2]}{1+2M_R \textcolor{magenta}{k }\bar{a} [1+(s-1)^2]} 
\\
-   2  {\mu}\frac{1}{  \textcolor{magenta}{k}} \frac{M_I(\bar{a} [1+(s-1)^2] )}{1+4M_R \textcolor{magenta}{k }\bar{a} [1+(s-1)^2]} \\
-  \frac{1}{1+   a^2 \textcolor{magenta}{k^2}} \frac{4 M_R
\bar{a} [1+(s-1)^2] a { \mu}  }{1+2M_R \textcolor{magenta}{k }\bar{a} [1+(s-1)^2]} \\
- 4 {\mu} \frac{M_R M_I  \bar{a}^{2} [1+(s-1)^2]^{2} }{1+4M_R \textcolor{magenta}{k }\bar{a} [1+(s-1)^2]}.
 \end{aligned}
\end{equation}}

\section{\label{comparison1ch}{Comparison with single-channel results}}

As an additional check, we compare our results with the previous work on single-channel elastic collisions \cite{Field:03,Guillon:09,Simoni:09,Bovino:11,Croft:17b,Mehta:18,Frye:19}. 

 To accomplish this, we start by finding the $Q_{11}$:
\begin{equation}
\begin{aligned}\label{q11}
Q_{11}= -i  S_{11}^\dagger \frac{d S_{11}}{d\textcolor{magenta}{E}}
=-i e^{-2 i \textcolor{magenta}{\xi_1}} \frac{1+ \frac{a_1^*}{\textcolor{magenta}{C_1^2}} }{1+\frac{1}{\textcolor{magenta}{C_1^2}} M_N^*}
\bigg(2 i \textcolor{magenta}{\xi_1}^\prime e^{2 i \textcolor{magenta}{\xi_1}} \frac{1+ \frac{a_1}{\textcolor{magenta}{C_1^2}}}{1+\frac{1}{\textcolor{magenta}{C_1^2}} M_N}  +e^{2 i \textcolor{magenta}{\xi_1}} \frac{ 2 \textcolor{magenta}{C_1}  \textcolor{magenta}{C_1^\prime} (M_N-a_1) }{(\textcolor{magenta}{C_1^2} + M_N)^2} \bigg)\\
=-i \frac{1+ \frac{a_1^*}{\textcolor{magenta}{C_1^2}} }{1+\frac{1}{\textcolor{magenta}{C_1^2}} M_N^*}
\bigg(2 i \textcolor{magenta}{\xi_1}^\prime  \frac{1+ \frac{a_1}{\textcolor{magenta}{C_1^2}}}{1+\frac{1}{\textcolor{magenta}{C_1^2}} M_N}  + \frac{ 2 \textcolor{magenta}{C_1}  \textcolor{magenta}{C_1^\prime} (M_N-a_1) }{(\textcolor{magenta}{C_1^2} + M_N)^2} \bigg).
 \end{aligned}
\end{equation}
When there is only one channel, $M_N=-iY_{\lambda_{11}}$ and $a_1=iY_{\lambda_{11}}$. Thus, Eq.~\eqref{q11} becomes
\begin{equation}
\begin{aligned}\label{q11}
Q_{11}=-i \frac{1- \frac{iY_{\lambda_{11}}}{\textcolor{magenta}{C_1^2}} }{1+\frac{iY_{\lambda_{11}}}{\textcolor{magenta}{C_1^2}} }
\bigg(2 i \textcolor{magenta}{\xi_1}^\prime  \frac{1+ \frac{iY_{\lambda_{11}}}{\textcolor{magenta}{C_1^2}}}{1-\frac{iY_{\lambda_{11}}}{\textcolor{magenta}{C_1^2}} }  + \frac{ 2 \textcolor{magenta}{C_1}  \textcolor{magenta}{C_1^\prime} (-2iY_{\lambda_{11}}) }{(\textcolor{magenta}{C_1^2} -iY_{\lambda_{11}})^2} \bigg),
 \end{aligned}
\end{equation}
which simplifies to 
\begin{equation}
\begin{aligned}\label{eqq}
Q_{11}= 
2  \textcolor{magenta}{\xi_1}^\prime   + \frac{ 2 \textcolor{magenta}{C_1}  \textcolor{magenta}{C_1^\prime} (-2Y_{\lambda_{11}}) }{
\textcolor{magenta}{C_1^4} +Y_{\lambda_{11}}^2} = 2  \textcolor{magenta}{\xi_1^\prime}  -   \frac{ 4 \textcolor{magenta}{C_1^\prime} Y_{\lambda_{11}} }{\textcolor{magenta}{C_1^3}( 1+ \frac{1}{\textcolor{magenta}{C_1^4}}Y_{\lambda_{11}}^2)} \\
 \end{aligned}
\end{equation}
In the s-wave regime, the MQDT parameters $\xi$ and $C_1^{-2}$ are defined by Eq.~\eqref{swave}. To find the time delay, we also require the derivatives of these parameters
  \begin{equation}
\begin{aligned} 
\frac{d \textcolor{magenta}{\xi}}{d\textcolor{magenta}{E}}=-\frac{ a { \mu} }{ (1+   a^2 \textcolor{magenta}{k^2})\textcolor{magenta}{k}},\\
\frac{d \textcolor{magenta}{C_1^{-2}}}{d\textcolor{magenta}{E}}={\mu}\textcolor{magenta}{k^{-1}}\bar{a} [1+(s-1)^2],\\
\frac{d \textcolor{magenta}{C_1}}{d\textcolor{magenta}{E}}={\mu} \frac{-1}{ 2 \textcolor{magenta}{k^{5/2}}\sqrt{\bar{a} [1+(s-1)^2]}}.
\end{aligned}
\end{equation}
Using Eq.~\eqref{eqq} and the derivatives of $C_1$ and $\xi^\prime$ in the s-wave regime we obtain
\begin{equation}
\begin{aligned}\label{eqq2}
Q_{11}=    -\frac{ 2 a { \mu} }{ (1+   a^2 \textcolor{magenta}{k^2})\textcolor{magenta}{k}} -  \frac{ 4 {\mu}\frac{-1}{ 2 \textcolor{magenta}{k^{5/2}}\sqrt{\bar{a} [1+(s-1)^2]}} (\textcolor{magenta}{k^{3/2}}\bar{a}^{3/2} [1+(s-1)^2]^{3/2} )  Y_{\lambda_{11}} }{(1+\textcolor{magenta}{k^2}\bar{a^2} [1+(s-1)^2]^2 Y_{\lambda_{11}}^2)}. \\
 \end{aligned}
\end{equation}
As we approach the limit $k\to 0$, Eq.~\eqref {eqq2} becomes
\begin{equation}
\begin{aligned}
Q_{11}=    -\frac{ 2 a { \mu} }{\textcolor{magenta}{k}} +   \frac{ 2 {\mu}  \bar{a} [1+(s-1)^2]  Y_{\lambda_{11}}}{\textcolor{magenta}{k}}.
 \end{aligned}
\end{equation}
The second term vanishes since we follow  Refs.~\cite{Jachymski:14,Chilcott:22} in taking $Y_{11}=0$, which implies $Y_{\lambda_{11}}=[Y_{11}^{-1}-\tan\lambda]^{-1}=0$.
The first term is identical to the single-channel time delay found in earlier work \cite{Field:03,Guillon:09,Simoni:09,Bovino:11,Croft:17b,Mehta:18,Frye:19}, thereby confirming our results.


Additionally, we will check the expression for the single-channel scattering phase shift. We start with the single-channel scattering  matrix  $S=e^{2 i \delta} $
\begin{equation}
\begin{aligned}
{S}
=e^{2 i \textbf{$\xi$}} \frac{1+ i \frac{1}{C_1^{2}} Y_{\lambda_{11}} }{1-i \frac{1}{C_1^{2}} Y_{\lambda_{11}}}.
\end{aligned}
\end{equation}
In terms of the scattering phase shift, we have
\begin{equation}
\begin{aligned}
e^{2i \delta} 
=e^{2 i \textbf{$\xi$}} \frac{1+ i \frac{1}{C_1^{2}} Y_{\lambda_{11}} }{1-i \frac{1}{C_1^{2}} Y_{\lambda_{11}}}, \\
2 i \delta=2 i \xi+\ln\bigg( \frac{1+ i \frac{1}{C_1^{2}} Y_{\lambda_{11}} }{1-i \frac{1}{C_1^{2}} Y_{\lambda_{11}}}\bigg),\\
 \delta= \xi- \frac{1}{2} i \ln\bigg( \frac{1+ i \frac{1}{C_1^{2}} Y_{\lambda_{11}} }{1-i \frac{1}{C_1^{2}} Y_{\lambda_{11}}}\bigg).\\
\end{aligned}
\end{equation}
The quantum defect parameter $C_1^{-2} \to \textcolor{magenta}{k}\bar{a} [1+(s-1)^2] \to 0$ in the s-wave regime ($k\to 0$),  so the expression for the phase shift becomes
\begin{equation}
\begin{aligned}
 \delta= \xi- \frac{1}{2} i \ln( 1) \Rightarrow
  \delta= \xi
\end{aligned}
\end{equation}
In the s-wave regime the expression becomes
\begin{equation}
\begin{aligned}
  \lim_{k\to0}\delta=   \lim_{k\to0} \atan(-ka) =-ka
\end{aligned}
\end{equation}
which is identical to the phase shift obtained previously \cite{Frye:19}.

\section{\label{two_channel_model}{Two-channel model}}

For our two-channel model, we choose a single-parameter expression for $\mathbf{Y}$  with $Y_{11}=Y_{22}=0$ and $Y_{12}\ne0$ as suggested by  Chilcott and Croft \cite{Chilcott:22}. To obtain $Q_{11}$ for the two-channel model, we expand Eq.~\eqref{q1111} by using the constants found in Eqs.~\eqref{consta} and \eqref{constb}.
\begin{equation}
\begin{aligned}
Q_{11}=\frac{1}{\bigg|1+\frac{1}{\textcolor{magenta}{C_1^2}} \bigg(\frac{1}{C_2^2}\frac{Y_{12}^2}{d^2  C_2^2}\frac{\frac{\tan\lambda_{1}}{d C_2^2}}{1+ \frac{\tan\lambda_{1}^2}{d^2 C_2^4}}- i \Big(\frac{\tan\lambda_{2}}{d }+
\frac{1}{C_2^2}\frac{Y_{12}^2}{d^2  C_2^2}\frac{1}{1+ \frac{\tan\lambda_{1}^2}{d^2 C_2^4}}\Big)\bigg)\bigg|^2}\\
 \cross \Bigg(
2  \textcolor{magenta}{\xi_1^\prime} \bigg|1- \frac{1}{\textcolor{magenta}{C_1^2}} \frac{1}{C_2^2}\frac{Y_{12}^2}{d^2  C_2^2}\frac{\frac{\tan\lambda_{1}}{d C_2^2}}{1+ \frac{\tan\lambda_{1}^2}{d^2 C_2^4}}- i \frac{1}{\textcolor{magenta}{C_1^2}}\Big(\frac{\tan\lambda_{2}}{d }+
\frac{1}{C_2^2}\frac{Y_{12}^2}{d^2  C_2^2}\frac{1}{1+ \frac{\tan\lambda_{1}^2}{d^2 C_2^4}}\Big)\bigg|^2 
+ \textcolor{magenta}{\xi_1^\prime}  \frac{1}{\textcolor{magenta}{C_1^2}}
4 \frac{1}{C_2^2}\frac{Y_{12}^2}{d^2  C_2^2}\frac{\frac{\tan\lambda_{1}}{d C_2^2}}{1+ \frac{\tan\lambda_{1}^2}{d^2 C_2^4}}\\
+  \frac{ 1}{|\textcolor{magenta}{C_1^2}+ \frac{1}{C_2^2}\frac{Y_{12}^2}{d^2  C_2^2}\frac{\frac{\tan\lambda_{1}}{d C_2^2}}{1+ \frac{\tan\lambda_{1}^2}{d^2 C_2^4}}- i (\frac{\tan\lambda_{2}}{d }+
\frac{1}{C_2^2}\frac{Y_{12}^2}{d^2  C_2^2}\frac{1}{1+ \frac{\tan\lambda_{1}^2}{d^2 C_2^4}})|^2} \\
\cross \bigg(4 \textcolor{magenta}{C_1}  \textcolor{magenta}{C_1^\prime}\Big(- \frac{\tan\lambda_{2}}{d }+
\frac{1}{C_2^2}\frac{Y_{12}^2}{d^2  C_2^2}\frac{1}{1+ \frac{\tan\lambda_{1}^2}{d^2 C_2^4}}+\frac{1}{\textcolor{magenta}{C_1^4}} \frac{1}{C_2^4}\frac{Y_{12}^4}{d^4 C_1^4 C_2^4}\frac{\frac{\tan\lambda_{1}^2}{d^2 C_2^4}}{(1+ \frac{\tan\lambda_{1}^2}{d^2 C_2^4})^2} (
\frac{1}{C_2^2}\frac{Y_{12}^2}{d^2  C_2^2}\frac{1}{1+ \frac{\tan\lambda_{1}^2}{d^2 C_2^4}}- \frac{\tan\lambda_{2}}{d })
\\+\frac{1}{\textcolor{magenta}{C_1^4}}\big(
\frac{1}{C_2^2}\frac{Y_{12}^2}{d^2  C_2^2}\frac{1}{1+ \frac{\tan\lambda_{1}^2}{d^2 C_2^4}}- \frac{\tan\lambda_{2}}{d }\big)^3\Big) \\
+  
 8\frac{ \textcolor{magenta}{C_1^\prime} }{\textcolor{magenta}{C_1}  }  \frac{1}{C_2^2}\frac{Y_{12}^2}{d^2  C_2^2}\frac{\frac{\tan\lambda_{1}}{d C_2^2}}{1+ \frac{\tan\lambda_{1}^2}{d^2 C_2^4}} (\frac{1}{C_2^2}\frac{Y_{12}^2}{d^2  C_2^2}\frac{1}{1+ \frac{\tan\lambda_{1}^2}{d^2 C_2^4}}- \frac{\tan\lambda_{2}}{d }) \bigg)\Bigg).
\end{aligned}
\end{equation}
This equation gives us the time delay of channel 1 ($Q_{11}$) in terms of the short-range MQDT parameter $Y_{12}$.

\section{\label{sec:Q11UM}{Derivation of the average time delay in the universal model}}

Equation (1) of Ref.~\cite{Idziaszek:10} gives the scattering length in terms of the diagonal S-matrix element in the entrance channel ($\alpha=1$ in our notation)
\begin{equation}
\begin{aligned}
\tilde{a}_{1}=\frac{1}{i k_{1}} \frac{1-S_{11}}{1+S_{11}}
\end{aligned}
\end{equation}
This can be rearranged to get the scattering matrix in term of the scattering length
\begin{equation}
\begin{aligned}
S_{11}= \frac{i+ k_{1} \tilde{a}_{1}}{i -k_{1} \tilde{a}_{1}}
\end{aligned}
\end{equation}
or 
\begin{equation}
\begin{aligned}
S_{11}= \frac{1-i k_{1} \tilde{a}_{1}}{1+i k_{1} \tilde{a}_{1}}
\end{aligned}
\end{equation}
Equation (11) of Ref.~\cite{Idziaszek:10} expresses the scattering length in terms of MQDT parameters 
\begin{equation}
\begin{aligned}
\tilde{a}_1=-\frac{1}{k}(\tan\xi-\frac{y C^{-2}(E)}{i+y\tan\lambda})
\end{aligned}
\end{equation}
In terms of the MQDT parameters, the S-matrix element takes the form 
\begin{equation}
\begin{aligned}
S_{11}= \frac{1+i (\tan\xi-\frac{y C^{-2}(E)}{i+y\tan\lambda})}{1-i  (\tan\xi-\frac{y C^{-2}(E)}{i+y\tan\lambda})}
\end{aligned}
\end{equation}
Since complex conjugation is distributive over division $\Big(\frac{z}{w}\Big)^*=\frac{z^*}{w^*}$, we have
\begin{equation}
\begin{aligned}
S_{11}^*= \frac{1-i (\tan\xi-\frac{y C^{-2}(E)}{-i+y\tan\lambda})}{1+i  (\tan\xi-\frac{y C^{-2}(E)}{-i+y\tan\lambda})}
\end{aligned}
\end{equation}
To find the energy derivative of $S_{11}$ we use the quotient rule $(\frac{u(x)}{v(x)})'=\frac{u’(x)v(x)-v’(x)u(x)}{v(x)^2}$
and the following relations
\begin{equation}
\begin{aligned}
\frac{d}{dx}(1 + i (g(x) - f(x)^{-2} a)) = i (g'(x) +2 a f’(x)f(x)^{-3}),\\
\frac{d}{dx}(1 - i (g(x) - f(x)^{-2}  a)) = -i (g'(x) + 2 a f’(x)f(x)^{-3}),
 \end{aligned}
\end{equation}
where $a=\frac{y}{-i+y\tan\lambda}$, $g(x)=\tan\xi$, and $f(x)=C(E)$.
We obtain 
 \begin{equation}
\begin{aligned}
\frac{ d S_{11}}{dE}= \frac{ i (g'(x) +2 a f’(x)f(x)^{-3}) (1-i  (\tan\xi-\frac{y C^{-2}(E)}{i+y\tan\lambda})+i (g'(x) + 2 a f’(x)f(x)^{-3})(1+i (\tan\xi-\frac{y C^{-2}(E)}{i+y\tan\lambda}))}{(1-i  (\tan\xi-\frac{y C^{-2}(E)}{i+y\tan\lambda}))^2}\\
= \frac{ i (g'(x) + 2 a f’(x)f(x)^{-3}) (1-i  (\tan\xi-\frac{y C^{-2}(E)}{i+y\tan\lambda})+1+i (\tan\xi-\frac{y C^{-2}(E)}{i+y\tan\lambda}))}{(1-i  (\tan\xi-\frac{y C^{-2}(E)}{i+y\tan\lambda}))^2}\\
= \frac{ 2 i (g'(x) + 2 a f’(x)f(x)^{-3})}{(1-i  (\tan\xi-\frac{y C^{-2}(E)}{i+y\tan\lambda}))^2}\\
\end{aligned}
\end{equation}
Substituting the values of $a$, $g(x)$, and $f(x)$ (see above), we find
 \begin{equation}
\begin{aligned}
\frac{ d S_{11}}{dE}
= \frac{ 2 i (\frac{d\tan\xi}{dE} + 2 \frac{y}{-i+y\tan\lambda} C’(E)C^{-3}(E))}{(1-i  (\tan\xi-\frac{y C^{-2}(E)}{i+y\tan\lambda}))^2}
\end{aligned}
\end{equation}

Now we can solve for the average time delay $Q_{11}$
\begin{equation}
\begin{aligned}
Q_{11}=- i S_{11}^\dagger \frac{d S_{11}}{dE} \\
= -i \frac{1-i (\tan\xi-\frac{y C^{-2}(E)}{-i+y\tan\lambda})}{1+i  (\tan\xi-\frac{y C^{-2}(E)}{-i+y\tan\lambda})} 
 \frac{ 2 i (\frac{d\tan\xi}{dE} + 2 \frac{y}{-i+y\tan\lambda} \frac{C’(E)}{C^{3}(E)})}{(1-i  (\tan\xi-\frac{y C^{-2}(E)}{i+y\tan\lambda}))^2}
\end{aligned}
\end{equation}

In the s-wave regime, where $\tan\xi=-ka$, $C^{-2}(E)=k \bar{a}[1+(s-1)^2]$, and $\tan\lambda=1-s$ \cite{Idziaszek:10}, the average time delay becomes 
\begin{equation}
\begin{aligned}
Q_{11}=-i S_{11}^\dagger \frac{d S_{11}}{dE}=\\
-i   \frac{1-i (-ka-\frac{y k \bar{a}[1+(s-1)^2]}{-i+y(1-s)})}{1+i  (-ka-\frac{y k \bar{a}[1+(s-1)^2]}{-i+y(1-s)})} \times
  \frac{ 2 i (\frac{d(-ka)}{dE} + 2 \frac{y}{-i+y(1-s)} \frac{d (k \bar{a}[1+(s-1)^2])^{-1/2}}{dE}(k \bar{a}[1+(s-1)^2])^{3/2})}{(1-i  (-ka-\frac{y k \bar{a}[1+(s-1)^2]}{i+y(1-s)}))^2}\\
=  -i   \frac{1+ ik a+ik\bar{a} y  \frac{[1+(s-1)^2]}{-i+y(1-s)}}{1 -ik a-ik\bar{a} y  \frac{[1+(s-1)^2]}{-i+y(1-s)}} \times 
  \frac{ 2 i (-a\frac{d(\sqrt{2{\mu} E})}{dE} + 2 \frac{y}{-i+y(1-s)} \frac{d ((\sqrt{2 {\mu} E})^{-1/2})}{dE}(k^{3/2} \bar{a}[1+(s-1)^2]))}{(1-i  (-ka-\frac{y k \bar{a}[1+(s-1)^2]}{i+y(1-s)}))^2}\\
 =     \frac{1+ ik a+ik\bar{a} y  \frac{[1+(s-1)^2]}{-i+y(1-s)}}{1-ik a-ik\bar{a} y  \frac{[1+(s-1)^2]}{-i+y(1-s)}} \times
  \frac{ -2  (-a\frac{\mu}{k} - 2 \frac{y}{-i+y(1-s)} (\frac{\mu}{2} k^{-5/2})(k^{3/2} \bar{a}[1+(s-1)^2]))}{(1-i  (-ka-\frac{y k \bar{a}[1+(s-1)^2]}{i+y(1-s)}))^2}\\
= \frac{1+ ik a+ik\bar{a} y  \frac{[1+(s-1)^2]}{-i+y(1-s)}}{1 -ik a-ik\bar{a} y  \frac{[1+(s-1)^2]}{-i+y(1-s)}} \frac{ 2(a+\frac{y  \bar{a}[1+(s-1)^2]}{i+y(1-s)}) \frac{\mu}{k}}{(1+aki+i\frac{y k \bar{a}[1+(s-1)^2]}{i+y(1-s)}  )^2},\\
\end{aligned}
\end{equation}

which we can write 
\begin{equation}\label{Q11_exp1}
\begin{aligned}
Q_{11}
= \frac{1+ ik a+ik\bar{a} y  \frac{[1+(s-1)^2]}{-i+y(1-s)}}{1 -ik a-ik\bar{a} y  \frac{[1+(s-1)^2]}{-i+y(1-s)}} \frac{ 2(a+\frac{y  \bar{a}[1+(s-1)^2]}{i+y(1-s)}) \frac{\mu}k}{(1+aki+i\frac{y k \bar{a}[1+(s-1)^2]}{i+y(1-s)}  )^2}\\
= \frac{1+ k \alpha}{1 -k \alpha} \frac{ 2 \beta \frac{\mu}{k}}{(1+i k\beta )^2}\\
= \frac{1+ k \alpha}{1 -k \alpha} \frac{ 2 \beta \mu}{k+2i k^2\beta-k^3 \beta^2 }
\end{aligned}
\end{equation}
where $\alpha=i a+i\bar{a} y  \frac{[1+(s-1)^2]}{-i+y(1-s)}$ and $\beta=a+\frac{y  \bar{a}[1+(s-1)^2]}{i+y(1-s)}$. Because $\beta$ and $\alpha$ are complex one cannot make the assumption 
\begin{equation}
\begin{aligned}
Q_{11} \approx    \frac{1+ k \alpha}{1 -k \alpha} \frac{2 \beta \mu}{k }
=2 \beta \mu  \frac{1+ k \alpha}{k -k^2 \alpha} 
\approx 2 \beta \mu  \frac{1+ k \alpha}{k } 
\end{aligned}
\end{equation}
when $k$ is small. Instead, we proceed  by separating Eq.~\eqref{Q11_exp1}  into the real and imaginary parts
\begin{equation}\label{lab}
\begin{aligned}
Q_{11}= \frac{1+ ik a+ik\bar{a} y  \frac{[1+(s-1)^2]}{-i+y(1-s)}}{1 -ik a-ik\bar{a} y  \frac{[1+(s-1)^2]}{-i+y(1-s)}} \frac{ 2(a+\frac{y  \bar{a}[1+(s-1)^2]}{i+y(1-s)}) \frac{\mu}{k}}{(1+aki+i\frac{y k \bar{a}[1+(s-1)^2]}{i+y(1-s)}  )^2}\\
= \frac{-i+y(1-s)+k a+ik a y(1-s) +ik\bar{a} y  [1+(s-1)^2]}{-i+y(1-s)-ka-i kay(1-s)-ik\bar{a} y  [1+(s-1)^2]}\\
\times
 \frac{ 2(ia+ay(1-s)+y  \bar{a}[1+(s-1)^2]) \frac{\mu}{k} (i+y(1-s))}{(i+y(1-s)-ka+i kay(1-s)+iy k \bar{a}[1+(s-1)^2]  )^2}. \\
\end{aligned}
\end{equation}
Now define a parameter 
\begin{equation}
\begin{aligned}
\beta=a y(1-s)+y  \bar{a}[1+(s-1)^2] \\
=a y(1-s)+y  \frac{a}{s}[1+s^2-2s+1]\\
=a y(1-s)+ay  \left[\frac{2}{s}+s-2\right]\\
=ay\left[\frac{2}{s}-1\right].
\end{aligned}
\end{equation}

With this definition, Eq. \eqref{lab} becomes
\begin{equation}
\begin{aligned}
Q_{11}= \frac{-i+y(1-s)+k a+ik \beta}{-i+y(1-s)-ka-i k\beta} \times
 \frac{ (ia+\beta) \frac{2\mu}{k}(i+y(1-s))}{(i+y(1-s)-ka+i k\beta  )^2}\\
\end{aligned}
\end{equation}
The denominator of the first term and the expression in parentheses in the denominator of the second term are complex conjugates of each other, and thus 
\begin{equation}
\begin{aligned}
Q_{11}=\frac{2\mu}{k} 
 \frac{ (ia+\beta)(i+y(1-s)) (-i+y(1-s)+k a+ik \beta)}{|i+y(1-s)-ka+i k\beta |^2 (i+y(1-s)-ka+i k\beta)}\\
 =\frac{2\mu}{k} 
 \frac{ (ia+\beta) (i+y(s-1))(-i+y(1-s)+k a+ik \beta)}{((1+k\beta)^2+(y(1-s)-ka)^2) (i+y(1-s)-ka+i k\beta)}
\end{aligned}
\end{equation}
To make the denominator real, we multiply the numerator and denominator by the denominator's complex conjugate 
\begin{equation}
\begin{aligned}
Q_{11}=\frac{2\mu}{k}
 \frac{ (ia+\beta)(i+y(s-1)) (-i+y(1-s)+k a+ik \beta)(-i+y(1-s)-ka-i k\beta)}{((1+k\beta)^2+(y(1-s)-ka)^2) |(i+y(1-s)-ka+i k\beta)|^2}\\
 =\frac{2\mu}{k}
\Bigg[\bigg(a^2 \beta k^2 s y - a^2 \beta k^2 y - 2 a s^2 y^2 + 4 a s y^2 - 2 a y^2 - a + \beta^3 k^2 s y + \beta^3 (-k^2) y + \beta s^3 y^3 - 3 \beta s^2 y^3 \\
+ 3 \beta s y^3 - \beta s y - \beta y^3 + \beta y\bigg)
+i\bigg(-a^3 k^2 s y + a^3 k^2 y - a \beta^2 k^2 s y + a \beta^2 k^2 y + a s^3 y^3 \\
- 3 a s^2 y^3 + 3 a s y^3 - a s y - a y^3 + a y + 2 \beta s^2 y^2 - 4 \beta s y^2 + 2 \beta y^2 + \beta\bigg) \Bigg]
\frac{1}{\bigg((1+k\beta)^2+(y(1-s)-ka)^2\bigg)^2}
\end{aligned}
\end{equation}

In the limit $k\to 0$  we can retain only the terms linear in $k$
\begin{equation}
\begin{aligned}\label{sec}
Q_{11}  =\frac{2\mu}{k}
 \frac{ (ia+\beta)(i+y(s-1)) (-i+y(1-s)+k a+ik \beta)(-i+y(1-s)-ka-i k\beta)}{((1+k\beta)^2+(y(1-s)-ka)^2) |(i+y(1-s)-ka+i k\beta)|^2}\\
 =\frac{2\mu}{k}
\Bigg(\bigg(a^2 \beta k^2 s y - a^2 \beta k^2 y - 2 a s^2 y^2 + 4 a s y^2 - 2 a y^2 - a + \beta^3 k^2 s y + \beta^3 (-k^2) y + \beta s^3 y^3 - 3 \beta s^2 y^3 \\
+ 3 \beta s y^3 - \beta s y - \beta y^3 + \beta y\bigg)
+i\bigg(-a^3 k^2 s y + a^3 k^2 y - a \beta^2 k^2 s y + a \beta^2 k^2 y + a s^3 y^3 \\
- 3 a s^2 y^3 + 3 a s y^3 - a s y - a y^3 + a y + 2 \beta s^2 y^2 - 4 \beta s y^2 + 2 \beta y^2 + \beta\bigg) \Bigg)\\
\times
\frac{1}{\bigg(1+2k\beta+k^2\beta^2+y^2(1-s)^2-2k a y(1-s)+k^2 a^2\bigg)^2}\\
=\frac{2\mu}{k}
\Bigg(\bigg( - 2 a s^2 y^2 + 4 a s y^2 - 2 a y^2 - a   + \beta s^3 y^3 - 3 \beta s^2 y^3 
+ 3 \beta s y^3 - \beta s y - \beta y^3 + \beta y\bigg)\\
+i\bigg(   a s^3 y^3 
- 3 a s^2 y^3 + 3 a s y^3 - a s y - a y^3 + a y + 2 \beta s^2 y^2 - 4 \beta s y^2 + 2 \beta y^2 + \beta\bigg) \Bigg)\\
\frac{1}{\bigg(1+y^2(1-s)^2+(2\beta-2ay(1-s))k\bigg)^2}\\
 =\frac{2{\mu}}{\textcolor{magenta}{k}}
\Bigg(\bigg( - 2 a s^2 y^2 + 4 a s y^2 - 2 a y^2 - a   + \beta s^3 y^3 - 3 \beta s^2 y^3 
+ 3 \beta s y^3 - \beta s y - \beta y^3 + \beta y\bigg)\\
+i\bigg(   a s^3 y^3 
- 3 a s^2 y^3 + 3 a s y^3 - a s y - a y^3 + a y + 2 \beta s^2 y^2 - 4 \beta s y^2 + 2 \beta y^2 + \beta\bigg) \Bigg)\\
 \times
\frac{1}{\bigg(1+y^2(1-s)^2\bigg)^2+2\bigg((1+y^2(1-s)^2)(2\beta-2ay(1-s))\bigg)\textcolor{magenta}{k}}\\
\end{aligned}
\end{equation}
In the limit of $k\to 0$, the tem $\left(1+(1-s)^2\right)^2+2\left((1+(1-s)^2)(2\beta-2a(1-s))\right)\textcolor{magenta}{k}$ goes over to $\left(1+(1-s)^2\right)^2$ and the expression for $Q_{11}$ becomes
\begin{equation}
\begin{aligned}\label{sec}
Q_{11} 
 =\frac{2\mu}{\textcolor{magenta}{k}}
\Bigg(\bigg( - 2 a s^2 y^2 + 4 a s y^2 - 2 a y^2 - a   + \beta s^3 y^3 - 3 \beta s^2 y^3 
+ 3 \beta s y^3 - \beta s y - \beta y^3 + \beta y\bigg)\\
+i\bigg(   a s^3 y^3 
- 3 a s^2 y^3 + 3 a s y^3 - a s y - a y^3 + a y + 2 \beta s^2 y^2 - 4 \beta s y^2 + 2 \beta y^2 + \beta\bigg) \Bigg)\\
\times
\frac{1}{\bigg(1+y^2(1-s)^2\bigg)^2}
\end{aligned}
\end{equation}

In the absence of inelastic loss ($y\to0$) this expression takes the form
\begin{equation}\label{lllll}
\begin{aligned}
\lim_{y\to0}Q_{11}  
 =\frac{2{\mu}}{\textcolor{magenta}{k}}
(- a+i  \beta)\\
\end{aligned}
\end{equation}
Recalling that  $\beta=ay[{2}/{s}-1]$, we have $\lim_{y\to0}\beta=0$ and  Eq.~\eqref{lllll} simplifies as
\begin{equation}
\begin{aligned}
\lim_{y\to0}Q_{11}  
 =-\frac{2a{\mu}}{\textcolor{magenta}{k}},\\
\end{aligned}
\end{equation}
which  is identical to the previous single-channel results \cite{Field:03,Guillon:09,Simoni:09,Bovino:11,Croft:17b,Mehta:18,Frye:19}. 

In the universal limit ($y\to1$), Eq.~\eqref{sec} becomes
\begin{equation}
\begin{aligned}
Q_{11} 
 =\frac{2\frac{\mu}{\hbar}}{\textcolor{magenta}{k}}
\Bigg(\bigg( - 2 a s^2  + 4 a s  - 2 a  - a   + \beta s^3  - 3 \beta s^2 
+ 3 \beta s  - \beta s  - \beta  + \beta \bigg)\\
+i\bigg(   a s^3  
- 3 a s^2  + 3 a s - a s  - a  + a  + 2 \beta s^2  - 4 \beta s  + 2 \beta  + \beta\bigg) \Bigg)\\
\times
\frac{1}{\bigg(1+(1-s)^2\bigg)^2+2\bigg((1+(1-s)^2)(2\beta-2a(1-s))\bigg)\textcolor{magenta}{k}}
\end{aligned}
\end{equation}
As noted above, in the limit of $k\to 0$, the term $\left(1+(1-s)^2\right)^2+2\left((1+(1-s)^2)(2\beta-2a(1-s))\right)\textcolor{magenta}{k}$ goes over to $\left(1+(1-s)^2\right)^2$, and we find
\begin{equation}
\begin{aligned}
Q_{11}  =\frac{2{\mu}}{\textcolor{magenta}{k}}
\Bigg(\bigg( - 2 a s^2  + 4 a s  - 3 a    + \beta s^3  - 3 \beta s^2 
+ 2 \beta s     \bigg)\\
+i\bigg(   a s^3  
- 3 a s^2  + 2 a s  + 2 \beta s^2  - 4 \beta s  + 3 \beta \bigg) \Bigg)
\times
\frac{1}{\bigg(1+(1-s)^2\bigg)^2}\\
\end{aligned}
\end{equation}
Recall that   $\beta=ay[{2}/{s}-1]$, so in this expression  $\beta=a[{2}/{s}-1]$. 

\subsection{Universal  time delay  for $p$-wave scattering}
For completeness,  here we provide a derivation of the average time delay in the $p$-wave regime ($l=1$).
We follow a similar procedure as outlined above  for the s-wave regime. The $p$-wave quantum defect parameters are given by \cite{Idziaszek:10} 
\begin{equation}
\begin{aligned}
C^{-2}=2.128 \textcolor{magenta}{k^3}\bar{a}^3 \frac{1+(s-1)^2}{(s-2)^2},\\
\tan\xi=2.128  \textcolor{magenta}{k^3}  \bar{a}^3 \frac{s-1}{s-2},\\
\tan\lambda=\frac{s}{s-2}.
 \end{aligned}
\end{equation}
The $p$-wave scattering length may be written as
\begin{equation}
\begin{aligned}
\tilde{a}&=-\frac{1}{k}(\tan\xi-\frac{y C^{-2}(E)}{i+y\tan\lambda})\\
&=-\frac{1}{ \textcolor{magenta}{k}}(2.128  \textcolor{magenta}{k^3}  \bar{a}^3 \frac{s-1}{s-2}-\frac{y  2.128 \textcolor{magenta}{k^3}\bar{a}^3 \frac{1+(s-1)^2}{(s-2)^2}}{i+y \frac{s}{s-2}})\\
&=-( 2\bar{a}_1 \textcolor{magenta}{k^2}  \bar{a}^2 \frac{s-1}{s-2}-\frac{y 2 \bar{a}_1 \textcolor{magenta}{k^2}\bar{a}^2 \frac{1+(s-1)^2}{(s-2)^2}}{i+y \frac{s}{s-2}})\\
&=-2 \bar{a}_1 \textcolor{magenta}{k^2}\bar{a}^2\bigg(  \frac{s-1}{s-2}-\frac{y  \frac{1+(s-1)^2}{(s-2)}}{i(s-2)+y s}\bigg)\\
&=-2 \bar{a}_1 \textcolor{magenta}{k^2}\bar{a}^2\bigg( \frac{(i(s-1)+y s \frac{s-1}{s-2} )- y  \frac{1+(s-1)^2}{(s-2)}}{i(s-2)+y s}\bigg).
\end{aligned}
\end{equation}
Pulling out the common factor $y$
\begin{equation}
\begin{aligned}
\tilde{a}=-2 \bar{a}_1 \textcolor{magenta}{k^2}\bar{a}^2\bigg( \frac{(i(s-1)+ y  \frac{s(s-1)-1-(s^2-2s+1)}{(s-2)}}{i(s-2)+y s}\bigg)\\
=-2 \bar{a}_1 \textcolor{magenta}{k^2}\bar{a}^2\bigg( \frac{(i(s-1)+ y  \frac{s^2-s-1-(s^2-2s+1)}{(s-2)}}{i(s-2)+y s}\bigg)
\end{aligned}
\end{equation}
and simplifying, we obtain
\begin{equation}
\begin{aligned}
\tilde{a}=-2 \bar{a}_1 \textcolor{magenta}{k^2}\bar{a}^2\bigg( \frac{i(s-1)+ y  \frac{s-2}{(s-2)}}{i(s-2)+y s}\bigg)\\
=-2 \bar{a}_1 \textcolor{magenta}{k^2}\bar{a}^2\bigg( \frac{i(s-1)+ y  }{i(s-2)+y s}\bigg).
\end{aligned}
\end{equation}
This scattering length is identical to the one obtained by Idziaszek and Julienne (see Eq. (19) of Ref.~\cite{Idziaszek:10}).
Thus, can write the $p$-wave S-matrix element in terms of MQDT parameters as
\begin{equation}
\begin{aligned}
S_{11}= \frac{1+  i2 \bar{a}_1 \textcolor{magenta}{k^3}\bar{a}^2\bigg( \frac{i(s-1)+ y  }{i(s-2)+y s}\bigg)}{1-i 2 \bar{a}_1 \textcolor{magenta}{k^3}\bar{a}^2\bigg( \frac{i(s-1)+ y  }{i(s-2)+y s}\bigg)}.
\end{aligned}
\end{equation}

To obtain the average time delay, we need the energy derivative of $S_{11}$ and its complex conjugate.
The energy derivative of $S_{11}$ can be found using the identity (where $f(x)$ is a differentiable function and $a$ is a constant)
\begin{equation}
\begin{aligned}
\frac{d}{dx}\frac{( (1 +i a f(x)^3)}{(1 -i a f(x)^3)} =-\frac{ 6 i a f(x)^2 f’(x)}{(1-ia f(x)^3 )^2},
\end{aligned}
\end{equation}
which gives
\begin{equation}
\begin{aligned}
\frac{d S_{11}}{dE}=-\frac{12i   \mu \bar{a}_1 \bar{a}^2\bigg( \frac{i(s-1)+ y  }{i(s-2)+y s}\bigg) \textcolor{magenta}{k} }{(1-i2 \bar{a}_1 \bar{a}^2\bigg( \frac{i(s-1)+ y  }{i(s-2)+y s}\bigg) \textcolor{magenta}{k^3} )^2}
\end{aligned}
\end{equation}

To obtain the complex conjugate of $S_{11}$, we use the fact that complex conjugation is distributive over division, i.e.,  $\Big(\frac{z}{w}\Big)^*=\frac{z^*}{w^*}$:
\begin{equation}
\begin{aligned}
S_{11}^*= \frac{\bigg(1+ i 2 \bar{a}_1 \textcolor{magenta}{k^3}\bar{a}^2\bigg( \frac{i(s-1)+ y  }{i(s-2)+y s}\bigg)\bigg)^*}{\bigg(1 -i 2 \bar{a}_1 \textcolor{magenta}{k^3}\bar{a}^2\bigg( \frac{i(s-1)+ y  }{i(s-2)+y s}\bigg)\bigg)^*}= \frac{1- i 2 \bar{a}_1 \textcolor{magenta}{k^3}\bar{a}^2\bigg( \frac{-i(s-1)+ y  }{-i(s-2)+y s}\bigg)}{1+i 2 \bar{a}_1 \textcolor{magenta}{k^3}\bar{a}^2\bigg( \frac{-i(s-1)+ y  }{-i(s-2)+y s}\bigg)}
\end{aligned}
\end{equation}

We are now able to obtain the average time delay as
\begin{equation}
\begin{aligned}
Q_{11}=- i S_{11}^\dagger \frac{d S_{11}}{dE}\\
=- \frac{12   \mu \bar{a}_1 \bar{a}^2\bigg( \frac{i(s-1)+ y  }{i(s-2)+y s}\bigg) \textcolor{magenta}{k} }{(1-i2 \bar{a}_1 \bar{a}^2\bigg( \frac{i(s-1)+ y  }{i(s-2)+y s}\bigg) \textcolor{magenta}{k^3} )^2} \frac{1- i 2 \bar{a}_1 \textcolor{magenta}{k^3}\bar{a}^2\bigg( \frac{-i(s-1)+ y  }{-i(s-2)+y s}\bigg)}{1+i 2 \bar{a}_1 \textcolor{magenta}{k^3}\bar{a}^2\bigg( \frac{-i(s-1)+ y  }{-i(s-2)+y s}\bigg)}\\
\\
=\frac{- 12   \mu \bar{a}_1 \bar{a}^2 \textcolor{magenta}{k}((s-1) - i y) (s y + i (s-2)) (-i 2 \bar{a}_1 \textcolor{magenta}{k^3}\bar{a}^2(y - i (s-1)) - i (s-2) + s y)}{(2 \bar{a}_1 \textcolor{magenta}{k^3}\bar{a}^2(y - i (s-1)) - (s-2) - i s y) (-i 2 \bar{a}_1 \textcolor{magenta}{k^3}\bar{a}^2(y + i (s-1)) + i (s-2) + s y)^2}
\end{aligned}
\end{equation}

Introducing a new variable $\tilde{k}=2 \bar{a}_1 \textcolor{magenta}{k^3}\bar{a}^2$, we obtain
\begin{equation}
\begin{aligned}
Q_{11}&=- i S_{11}^\dagger \frac{d S_{11}}{dE}\\
=&-\frac{6 \mu }{k^2} 
\frac{\Big(\tilde{k}(s-2)[(s-1)^2+y^2]+y^3s^2+(s-2)^2y\Big)+i\Big([(s-2)+y^2s^2](s-1)+\tilde{k}ys[(s-1)^2-y^2]\Big)\bigg)\Bigg)}{i\Big((s-2)-\tilde{k}y\Big)+\Big([\tilde{k}y-(s-2)^2]^2+[ys+\tilde{k}(s-1)]^2+ys+\tilde{k}(s-1)\Big)}
\end{aligned}
\end{equation}
which can be rearranged to give the final result for the average time delay in $p$-wave scattering
\begin{equation}
\begin{aligned}
Q_{11} = -\frac{6 \mu }{k^2} \Bigg( \bigg(\Big([\tilde{k}y-(s-2)^2]^2+[ys+\tilde{k}(s-1)]^2+ys+\tilde{k}(s-1)\Big)\Big(\tilde{k}(s-2)[(s-1)^2+y^2]+y^3s^2+(s-2)^2y\Big)\\
-\Big((s-2)-\tilde{k}y\Big)\Big([(s-2)+y^2s^2](s-1)+\tilde{k}ys[(s-1)^2-y^2]\Big)\bigg)
+i\bigg(\Big((s-2)-\tilde{k}y\Big)\Big(\tilde{k}(s-2)[(s-1)^2+y^2]+y^3s^2+(s-2)^2y\Big)\\
\times
\Big([\tilde{k}y-(s-2)^2]^2+[ys+\tilde{k}(s-1)]^2+ys+\tilde{k}(s-1)\Big)\Big([(s-2)+y^2s^2](s-1)+\tilde{k}ys[(s-1)^2-y^2]\Big)\bigg)\Bigg)\\
\times
\frac{1}{\Big((s-2)-\tilde{k}y\Big)^2+\Big([\tilde{k}y-(s-2)^2]^2+[ys+\tilde{k}(s-1)]^2+ys+\tilde{k}(s-1)\Big)^2}
\end{aligned}
\end{equation}

\bibliography{master_v2.bib}
\end{document}